\documentclass{sig-alternate}
\usepackage{subfigure,graphicx}
\usepackage{multirow,rotating}
\usepackage[ruled,vlined, linesnumbered,shortend]{algorithm2e}
\newcommand{\up}{\vspace*{-0.0cm}}
\newcommand{\hup}{\vspace*{-0.0cm}}
\newcommand{\down}{\vspace*{0.2cm}}

\begin{document}
%

%
\title{Exploring Social Influence for Recommendation - A Probabilistic Generative Model Approach}

\numberofauthors{1}
\author{\alignauthor Mao Ye \hspace{10pt} Xingjie Liu \hspace{10pt} Wang-Chien Lee \\
\affaddr{Department of Computer Science and Engineering, The Pennsylvania State University, PA, USA.} \\
\email{\{mxy177,xzl106,wlee\}@cse.psu.edu }
}

\maketitle

\begin{abstract}
In this paper,  we propose a probabilistic generative model, called \emph{unified model}, which naturally unifies the ideas of social influence, collaborative filtering and content-based methods for item recommendation. To address the issue of hidden social influence, we devise new algorithms to learn the model parameters of our proposal based on expectation maximization (EM). In addition to a single-machine version of our EM algorithm, we further devise a parallelized implementation on the Map-Reduce framework to process two large-scale datasets we collect. Moreover, we show that the social influence obtained from our generative models 
can be used for group recommendation. Finally, we conduct comprehensive experiments using the datasets crawled from last.fm and whrrl.com to validate our ideas. Experimental results show that the generative models with social influence significantly outperform those without incorporating social influence. The unified generative model proposed in this paper obtains the best performance. Moreover, our study on social influence finds that users in whrrl.com are more likely to get influenced by friends than those in last.fm. The experimental results also confirm that our social influence based group recommendation algorithm outperforms the state-of-the-art algorithms for group recommendation.
\end{abstract} 

\up\hup

\up\hup

\up\hup

\down\down

\section{Introduction}\label{sec:intro}
As an indispensable type of information filtering techniques, recommendation systems have attracted a lot of attention in the past decade and have been successfully deployed in many e-commerce websites, such as Amazon and Netflix. \emph{Collaborative filtering} (CF) and \emph{content-based} techniques are two widely adopted approaches for recommendation systems~\cite{AdomaviciusT05TKDE}. Collaborative filtering~\cite{BreeseHK98UAI,Goldberg92CACM,HerlockerKBR99SIGIR,SarwarKKR01WWW,Wang2006SIGIR}  recommends items for a given user by referencing item ratings from other similar users, while content-based techniques~\cite{Mooney00DL} make recommendations by matching a user's personal interests (or profiles) with item content (e.g., item description or tags). Some research works have also discussed approaches that integrate both techniques for item recommendation~\cite{PopesculUPL01UAI,ZhengZXY10WWW}.
However, no emphasis has been placed explicitly on users' \emph{social influence} in these works. In our real life, we usually turn to our friends for recommendations of books, movies or restaurants. As evident by the dramatic expansion of social media and social networking systems, social influence from friends presents new opportunities for recommendation systems but also bring many great challenges. In this paper, we aim to take social influence among users, along with user profile, user preference and item content, into the design of recommendation systems.

To meet users' social demands, \cite{JamaliE09KDD,KonstasSJ09SIGIR,MaKL09Sigir,MaLK09recsys,Ma11WSDM} show that social influence is beneficial for item recommendations. The idea behind is that a user's friends may share common interests with the user, and have influence on the user's decisions. To incorporate the social influence to the recommendation system, \cite{JamaliE09KDD,KonstasSJ09SIGIR} employ the random walk approach~\cite{TongFP06ICDM} to incorporate user's social network for item recommendation. On the other hand, model-based systems were also been extended to include social influence~\cite{MaKL09Sigir,MaLK09recsys,Ma11WSDM}. Assuming that trust intensities among a user and his friends are available, some prior works propose to integrate users' social trust network into their models through a linear combination~\cite{MaKL09Sigir,MaLK09recsys} or as a regularization term~\cite{Ma11WSDM}. However, most of the proposed methods either apply ad hoc heuristics to include social influence to their methods or assume quantified prior knowledge of social trusts which is handily available. There is a need to define comprehensible social influence, beyond random walking over the social network, to explicitly model and unveil the social influence from data available to the recommendation system.

Owing to the success of collaborative filtering and content-based recommendation ideas, in this paper we propose to incorporate social influence with these ideas in a unified fashion to design new recommendation systems. Through our design, we aim to demonstrate the importance and strength of social influence to recommendation services.
To our best knowledge, ideas for unifying social influence with collaborative filtering and content-based recommendation are unexplored and very challenging. In this paper, we adopt the \emph{probabilistic generative model} as a methodology to reach our goal. The basic idea behind probabilistic generative models is to ``mimic'' user behaviors in a process of decision making, e.g., deciding which restaurant to dine. While there exists a prior   study~\cite{PopesculUPL01UAI} on integrating collaborative filtering and item content into a probabilistic generative model for item recommendation, we want to point out that incorporating social influence into the probabilistic generative model is nontrivial. Notice that the data used for CF and content based techniques (i.e., the user-item accessing history, user profiles and item content) contain \emph{explicit} observations. Thus, the notions of user preferences, user profiles and item content can be easily modeled. On the other hand, social influence cannot be observed directly from the data (i.e., we only know who access which item but never know if this decision is influenced by other people), we aim to introduce a latent variable and develop algorithms to capture the social influence between friends in addition of the latent variable for user's topics.

The proposed probabilistic generative model is a latent class statistical mixture model. The model discovers (1) users' personal preference distribution over latent topics\footnote{The term topic, from topic models, represents a genre of items in this paper. Take movies as an example of the items, a topic could be action, thriller, romantic or even a latent genre that cannot be expressed literally.}; (2) an item generative distribution for each topic; and (3) a social influence distribution from friends for each user. The generative model aims to capture the process of human behaviors and/or reasonings for decision making. For example, a user ($u$) wants to choose a restaurant ($i$) for dinner. He may choose one based on his own tastes or turn to one of his friends ($f$) for help. In the case that $u$ wants to choose the restaurant without any influence from his friends (with a certain probability), he chooses a topic according to his personal preference distribution. Then the selected topic in turn ``generates'' an item $i$ following on the topic's item generative distribution. In the case that social influence from a friend $f$ is effective, $f$ would generate an item following $f$'s preference distributions similarly. Thus, this model simulates the process that how $u$ picks the item $i$, including how a friend $f$ influences $u$'s decision. 

As mentioned, both users' preferences and social influence among friends are latent variables. Thus, there is a need to devise new learning algorithms to estimate the model parameters. In this paper, we address this issue by devising a new model learning algorithm based on the idea of \emph{expectation maximization} (EM). Moreover, due to the large volume of social network datasets and the excessive computational cost incurred in learning the generative model parameters, we devise a parallel algorithm under the Map-Reduce framework in addition to a single-thread algorithm, to process the large-scale datasets we collect. Finally, to demonstrate the flexibility and applicability of our ideas to other recommendation services that may utilize social influence, we adapt our probabilistic generative model and develop an algorithm to support \emph{group recommendations}. The primary contributions made in our research are summarized as follows.

\begin{itemize}
\up
\item We argue that social influence is important for item recommendations and devise probabilistic generative models that explicitly quantify and incorporate social influence from friends to a user in the recommendation process.
\up
\item We provide model learning methods (based on EM algorithms) to learn the model parameters from common user-item pairs. We implement the algorithms on single machine and parallel processing platform (based on the Map-Reduce framework~\cite{Jeffrey08MapReduce}) to  efficiently process large-scale data.
\up
\item In addition to support item recommendation for individual users, we demonstrate that the quantified social influence parameter is essential for supporting group recommendations. Owing to the advantages of social influence learned in our model, the proposed social-influence-based   group recommendation algorithms  significantly outperforms conventional aggregation-based allgorithms.
\up
\item  We conduct a comprehensive performance evaluation on two real datasets crawled from last.fm and whrrl.com.  Experimental results show that our proposal to incorporate social influence into generative models for item recommendation techniques are very effective. 
    The experimental results for group recommendation also confirm that the good estimation of social influence in our generative model is beneficial for group recommendation.
\end{itemize}

\up
The remainder of this paper is organized as follows. Section~\ref{sec:preliminary} summarizes the related work and provide some background on probabilistic generative models. Section~\ref{sec:social} introduces the design of our generative model which combines collaborative filtering and social influence into recommendation process. 
Section~\ref{sec:mapreduce} discusses how the EM algorithm is implemented on a single machine and on the Map-Reduce framework. Section~\ref{sec:extend} demonstrates how to incorporate social influence, in addition of collaborative filtering and item content, into the probabilistic generative model. Section~\ref{sec:group} reviews some previous group recommendation methods and proposes a new group recommendation method using the social influence obtained from our model. Section~\ref{sec:perform} shows the result of an empirical evaluation of our proposal using two real datasets. Finally, Section~\ref{sec:conclusion} concludes the paper. 

\up
\section{Preliminary}\label{sec:preliminary}
In this section, we introduce some related works, including recommendation systems, recommendation in social networks and group recommendation. 
Then, we provide the background about how to utilize probabilistic generative model for item recommendations.

\subsection{Related Work}
\noindent{\textbf{Recommendation System}.} Item recommendation has been a crucial service for many e-commerce and web services (e.g. netflix.com and amazon.com). The goal is to recommend an accurate list of items that the targeted user may be interested in. Collaborative filtering and content-based techniques are two widely adopted approaches for recommendation systems~\cite{AdomaviciusT05TKDE}. Both of them discover users' personal interests and utilize these interests to find relevant items. Collaborative filtering techniques ~\cite{BreeseHK98UAI,Goldberg92CACM,HerlockerKBR99SIGIR,SarwarKKR01WWW,Wang2006SIGIR,ZhengZXY10WWW} automatically predict relevant items for a given user by referencing item rating information from other similar users.  Content-based techniques~\cite{Mooney00DL} make recommendations by matching a user's personal interests (or profiles) to descriptive item information.
Recommendation systems using pure collaborative filtering approaches tend to fail when little knowledge about the user is known or when no one has similar interests with the user. For example, if a user has little item rating/selection history or his interests are rare compared to others, the item rating/selection history of other users cannot help. Although content-based methods is able to cope with the issue of lacking knowledge, it fails to account for community endorsement. For example, even though we know a user is interested in Chinese restaurants, content-based methods may possibly recommend a bad Chinese restaurant to him due to the lack of consideration in users' group consensus. As a result, there has been a continuous research interests and effort in combining the advantages of both collaborative filtering and content-based methods~\cite{BasuHC98AAAI,PopesculUPL01UAI,Basilico04ICML,KimLPKK06JIIS}. Our proposal in this work not only is able to naturally integrate the ideas behind collaborative filtering and content-based methods but also incorporate social influence into the recommendation process.

\noindent{\textbf{Social Recommendation}.} Under the context of social networks, social friendship is shown to be beneficial for recommendation ~\cite{Ma11WSDM,MaLK09recsys,MaKL09Sigir,JamaliE09KDD,KonstasSJ09SIGIR,Mao11SIGIR,Mao10GIS}. However, prior works in this area are mostly based on ad hoc heuristics. How a user is influenced by friends in the item selection process remains vague. For example, \cite{Mao11SIGIR} linearly combines social influence with conventional collaborative filtering; \cite{JamaliE09KDD,KonstasSJ09SIGIR} employ the random walk~\cite{TongFP06ICDM} approach to incorporate social network information into the process of item recommendation; while \cite{Ma11WSDM,MaLK09recsys,MaKL09Sigir} explores social friendship via matrix factorization technique,  where social influence is integrated by simple linear combination ~\cite{MaLK09recsys,MaKL09Sigir} or as a regularization term~\cite{Ma11WSDM}.

In this paper, we propose to employ the probabilistic generative model as a methodology to integrate social influence with collaborative filtering and content-based methods for item recommendation. Our work is uniquely different from these previous works because we do not assume social influence is explicitly available. By leaning a quantitative parameter for social influence, we are able to obtain a better understanding of the social influence and improve the performance of recommendation systems. Moreover, the quantified social influence obtained in our model can support related applications such as group recommendation~\cite{BaltrunasMR10RecSys,BerkovskyF10RecSys} and viral marketing~\cite{RichardsonD02KDD,ChenWW10KDD,GoyalBL10WSDM}.

\noindent{\textbf{Group Recommendation}.} To explore how to utilize social influence for group recommendation, we provides an in-depth study and comparison on group recommendation techniques. Group recommendations have been designed for various domains such as web/news pages~\cite{Pizzutilo05WSEAS}, tourism~\cite{McCarthySCMSN06FLAIRS}, music~\cite{McCarthy98CSCW,Crossen02IUI}, and TV programs and movies~\cite{OConnor01,Yu06UMUAI}. In summary, two main approaches have been proposed for group recommendation~\cite{Jameson07AW}. The first one creates an aggregated profile for a group based on its group members and then makes recommendations based on the aggregated group profile~\cite{McCarthy98CSCW,Yu06UMUAI}. The second approach \emph{aggregates} the recommendation results from individual members into a single group recommendation list. In other words, recommendations (i.e., ranked item lists) for individual members are created independently and then aggregated into a joint group recommendation list~\cite{BaltrunasMR10RecSys}, where the aggregation functions could be based on average or least misery strategies~\cite{Masthoff04UMUAI}. Different from these proposed methods, our approach regenerates the process of how group members would \emph{express} their preferences and \emph{influence} other members to reach the final decision. Evaluation from real datasets demonstrates a significant improvement over the proposed method using social influence over the traditional methods.

\subsection{Background}
\begin{figure}[htl]
\centering
\includegraphics[height=0.35in]{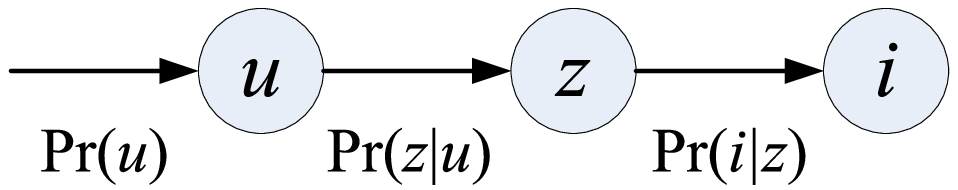}
\up
\caption{Probabilistic generative model- collaborative filtering}\label{fig:cf}
\up\hup
\end{figure}
The recommendation techniques we proposed in this paper are inspired by the probabilistic generative model developed for collaborative filtering in \cite{HofmannP99IJCAI}. 
Let $U=\{u_1,u_2,\cdots, u_N\}$ and $I=\{i_1,i_2,\cdots, i_M\}$ be the user set and item set, respectively. A latent topic set $Z=\{z_1,z_2,\cdots,z_K\}$ is assumed to capture latent user interests and item profiles. In the context of item recommendation, an event of a user $u\in U$ accessing an item $i\in I$ is considered to be associated with one of the latent topic variables $z\in Z$. Conceptually, as shown in Figure~\ref{fig:cf}, user $u$ chooses a topic $z \in Z$ according to his interest distributions, and in turn the topic $z$ probabilistically ``generates'' an item $i$ according to the distribution of items associated with $z$. Under this model, users are assumed to be independent of items given the chosen topic. The joint probability distribution over user $u$, topic $z$ and item $i$ can be written as \up
\begin{equation*}
\Pr(u,z,i) = \Pr(u)\Pr(z|u)\Pr(i|z),
\up
\end{equation*}
An equivalent specification of the joint probability distribution that treats users and items symmetrically is
\begin{equation*}
\Pr(u,z,i) = \Pr(z)\Pr(u|z)\Pr(i|z)
\up
\end{equation*}
Since we are only interested in how likely a user $u$ chooses an item $i$, the joint distribution over $u$ and item $i$ is
\begin{equation}\label{eq:cf}
\Pr(u,i) = \sum_{z\in Z} \Pr(u,z,i) = \sum_{z\in Z} \Pr(z)\Pr(u|z)\Pr(i|z)
\up
\end{equation}

This model has a set of parameters $\Pr(z)$, $\Pr(u|z)$ and $\Pr(i|z)$ for all $z\in Z, u\in U, i\in I$, which for simplicity is represented as $\theta$. In~\cite{HofmannP99IJCAI}, the user-item concurrence history $H = \{\langle u,i\rangle\}$, which contains all the observed user-item, is used to learn the model parameters $\theta$. One way to learn $\theta$ is to maximize the log-likelihood of history data which is: \up
\begin{equation}\label{eq:loglikelihood-cf}
\mathcal{L}(\theta) = \sum_{\langle u,i\rangle \in H} \log(\Pr(u,i|\theta)),
\up
\end{equation}
where each $\Pr(u,i|\theta)$ can be found using model parameters as in Equation~(\ref{eq:cf}).

After model parameters are inferred, items can be ranked for a given user according to $\Pr(i|u)$, which refers to the probability that the user $u$ selects the item $i$. $\Pr(i|u)$ can be computed as \up
\begin{equation}\label{eq:cf_rec}
\Pr(i|u) = \frac{\Pr(u,i)}{\Pr(u)} \propto \Pr(u,i)
\up
\end{equation}
Since most recommendation systems only focus on recommending \emph{new} items (items not presented in $H$ for a particular user $u$), items with the higher $\Pr(i|u)$ and not accessed by $u$ are good recommendations.

The probabilistic generative model described above is based on the ideas of collaborative filtering. Although \cite{PopesculUPL01UAI} has extended the model to integrate item contents as an additional component, social influence has not been considered yet. Moreover, as to be shown later, incorporating social influence into the generative model is fundamentally more challenging than integrating item contents into the model because item contents are \emph{observable} from the training data, while the social influence is a hidden factor not directly observable. In this paper, we demonstrate how to integrate social influence into this model and introduce our approach to infer model parameters.

\up
\section{Social Influence in Action}\label{sec:social}
In this section, we introduce our approach to incorporate social influence in a new probabilistic generative model for item recommendation. While the ultimate goal of our study is to unify the ideas of social influence, collaborative filtering and content-based methods as a model for item recommendation. For simplicity, here we first discuss how to integrate collaborative filtering and social influence into a probabilistic generative model. We will introduce the complete model (including collaborative filtering, item content and social influence) later in Section~\ref{sec:extend}. \up

\begin{figure}[thl]
\centering
\includegraphics[height=0.35in]{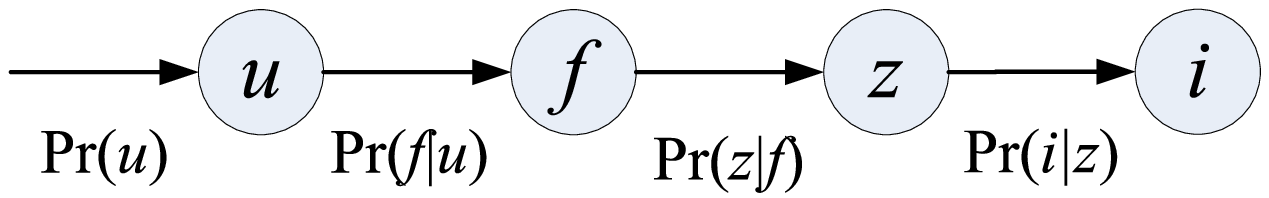}
\up
\caption{Probabilistic generative model combining social influence and collaborative filtering}\label{fig:cf-social}
\up
\end{figure}
Here we propose a new probabilistic generative model that describes the process of how a user selects an item by taking into account the user's own preferences and the social influence from her friends. Let $F(u) \subseteq U$ denote the friend list of a user $u$. The social influence introduced in this model aims to capture the scenario that one of $u$'s friends ($f \in F(u)$) has contributed his opinions in the item selection process. Here, for simplicity, we assume $u$ is a special friend of himself (i.e., $u \in F(u)$). Therefore, our model can be depicted as in Figure~\ref{fig:cf-social}. As shown, a user $u$ would first picks a friend (including himself) $f\in F(u)$ to make the item selection. If the picked friend $f$ happens to be himself $(f=u)$, $u$ is not influenced by someone else in this selection. Nevertheless, if the picked friend $f$ is not $u$, $u$ is influenced by $f$ at this time and thus the selected item follows $f$'s interests rather than $u$'s own tastes. In this model, we define a parameter \emph{social influence distribution} $\Pr(f|u)$ as the probability for $u$ to be influenced by a friend $f$.\footnote{We will demonstrate later that this parameter is very useful not only for item recommendation to individual users but also
for group recommendation.}
After $f$ is chosen based on $\Pr(f|u)$, $f$ randomly chooses a topic $z$ according to his interests, and then the topic generates an item $i$ according to the topic's item distribution.

The joint probability distribution over users, friends, topics and items is as below.
\begin{equation}\label{eq:joint} \up
\Pr(u,f,z,i) = \Pr(u)\Pr(f|u)\Pr(z|f)\Pr(i|z)
\end{equation}
where $u\in U$, $i \in I$, $f\in F(u)$ and $z\in Z$. 

The key observations from this model are 1) $u, z$ and $i$ are independently conditioned on $f$, and 2) $u,f$ and $i$ are independently conditioned on $z$. 
Because we intend to model the item selection probability in terms of social influence and topics (two latent parameters in our model), we transform Equation~(\ref{eq:joint}) into the following form: 
\begin{equation}\label{eqn:zifu}
\begin{split}
\Pr(u,f,z,i) & = \Pr(u|f,z,i) \Pr(f,z,i) = \Pr(u|f) \Pr(f,z,i)\\
 & = \Pr(z)\Pr(u|f)\Pr(f|z)\Pr(i|z)
\end{split}
\up\up\up\up
\end{equation}
Thus, the joint distribution over users and items is:\up
\begin{equation}\label{eq:cf-social-model}
\Pr(u,i) = \sum_{z\in Z}\sum_{f\in F(u)} \Pr(z)\Pr(u|f)\Pr(f|z)\Pr(i|z)
\up
\end{equation}

Different from the generative model in~\cite{HofmannP99IJCAI} (see Equation~(\ref{eq:cf})), 
the newly proposed model has two latent variables, namely the topic variable ($z$) and the social influence variable ($f$). Correspondingly, the model parameters $\theta$ now include $\{\Pr(z), \Pr(u|f), \Pr(f|z), \Pr(i|z)\}$. Note that the size of parameters is increased by $|U|\cdot|\overline{F(u)}|$ for social influence $\Pr(u|f)$. Notice that while the friend space could potentially be the entire user space, the averaged number of friends per user is limited.\footnote{In our collected real data sets, the averaged number of friends per user is less than 10.} This is very important because it ensures that our latent variable space is small enough and not to over-complicate the model. Moreover, the small latent parameter space yields high-quality parameter estimations even when the available history $H$ is not large.

In this study, we employ \emph{expected maximization (EM)} to learn model parameters from the user-item history $H$. However, the conventional expected maximization (EM) algorithm developed for single latent variable is not applicable for our model because we now have two latent variables, i.e., social influence and topics. To address this challenging issue, we have performed a detailed mathematical derivation to develop an new EM algorithm 
in order to infer the model parameters.\footnote{Due to space limit, we present the algorithm here but keep the EM algorithm derivation in the appendix.}

The derived EM algorithm iterates over the following steps: \up
\begin{enumerate}
  \item E-step: Computes posterior of the latent variables as $\Pr(z,f|u,i), \forall \langle u, i \rangle \in H$ using the model parameters of previous iteration. \up
  \item M-step: Computes new model parameters by maximizing the expected log-likelihood. \up
\end{enumerate}

In the E-step, we only need to use the previous iteration's model parameters to find $\Pr(z,f|u,i)$ as: \up
\begin{equation}\label{eq:estep}
\begin{split}
&\Pr(z,f|u,i) \\= &\frac{\Pr(z) \Pr(f|z)\Pr(u|f) \Pr(i|z)}{\sum_{z\in Z}\sum_{f\in F(u)} \Pr(z) \Pr(f|z)\Pr(u|f) \Pr(i|z)} \up
\end{split}
\end{equation}
Also note that we only need to compute the posteriors of those pairs presented in the history $H$ instead of all the possible user-item pairs, because the expectation to be maximized only weights on the observed user-item pairs.

The M-step shall find new model parameters to maximize the expected log-likelihood found in the E-step. According to our derivation, the new model parameters should be updated as
\begin{equation}\label{eq:prmstep}
\begin{split}
\Pr^+(z) & \propto \sum_{\langle u',i'\rangle \in H}\sum_{f' \in F(u')} \Pr(z,f'|u',i')\\
\Pr^+(u|f) & \propto \sum_{\langle u,i'\rangle \in H \wedge f\in F(u)}\sum_{z' \in Z} \Pr(z',f|u,i')\\
\Pr^+(f|z) & \propto \sum_{\langle u',i'\rangle \in H \wedge f\in F(u')} \Pr(z,f|u',i') \\
\Pr^+(i|z) & \propto \sum_{\langle u',i\rangle \in H}\sum_{f' \in F(u')} \Pr(z,f'|u',i)
\end{split}
\end{equation}

Equation~(\ref{eq:prmstep}) shows that for each parameter distributions, the new number should be chosen as normalized corresponding posterior sums. For example, $\Pr^+(f|z)$ is obtained by taking the sum of all the \emph{related} latent variable posteriors for the $f$ and $z$. Then, because of $\sum_{f\in U} \Pr^+(f|z) = 1$, we need to normalize the posterior sums with regarding to different $f$ to update the correct model parameters of $\Pr^+(f|z)$.

By repeating the E-step and M-step, the EM Algorithm improves the model parameters iteratively until they converge to a local log-likelihood maximum. The learned model parameters are used for item recommendations by ranking items for a given user according to
\begin{equation}
\Pr(i|u) \propto \sum_{f\in F(u),z \in Z}\Pr(u,f,z,i)
\end{equation}
which can be calculated by Equation~(\ref{eqn:zifu}).

\up
\section{Learning Algorithms}\label{sec:mapreduce}
In this section, we discuss how to implement the EM algorithm efficiently to learn the model parameters. Here we first present our algorithm for single machine. A challenge encountered in our initial research effort is that the EM algorithm, while fine tuned, is still slow due to excessive computation incurred in processing large-scale datasets. To overcome this challenge, we develop a parallel processing version of the EM algorithm on the Map-Reduce framework.
Through this effort, we demonstrate that our design of the EM algorithm can be elegantly decomposed for efficient parallel processing using Map-Reduce.

\subsection{Single Machine Algorithm}
We first show an implementation that efficiently realize our EM algorithm on a single machine. For simplicity, we only present one iteration of the E-step and the M-step, which aims is to approach the model parameters $\theta_{x+1}$ based on the current approximate value of parameters $\theta_x$.
\begin{algorithm}[htb]
\small
\KwIn{Data Set: ${H} = \{\langle u, i\rangle\}$, Model parameters: $\theta_x = \{\Pr(u|f),\Pr(f|z),\Pr(i|z),\Pr(z)\}$}
\KwOut{Next Parameters: $\theta_{x+1} = \{\Pr^+(u|f),\Pr^+(f|z),\Pr^+(i|z),\Pr^+(z)\}$}
\For{$\langle u, i \rangle \in H$}  {
    \For{$f \in F(u)$}  {
        \For{$z \in Z$} {
           Compute $\Pr(f,z|u,i)$\;\label{socialem:estep}
            $\Pr^{+}(f|z) \gets \Pr^{+}(f|z) + \Pr(f,z|u,i)$\;\label{socialem:mstep}
            $\Pr^{+}(i|z) \gets \Pr^{+}(i|z) + \Pr(f,z|u,i)$\;
            $\Pr^{+}(u|f) \gets \Pr^{+}(u|f) + \Pr(f,z|u,i)$\;
            $\Pr^{+}(z) \gets \Pr^{+}(z) + \Pr(f,z|u,i)$\;
        }
    }
}
Normalize $\Pr^+(z), \Pr^{+}(f|z), \Pr^{+}(i|z), \Pr^{+}(u|f)$\;\label{socialem:norm}
\Return{ $\theta_{x+1} = \{\Pr^+(u|f),\Pr^+(f|z),\Pr^+(i|z),\Pr^+(z)\}$}
\caption{\small{Social Influence EM Algorithm}}
\label{alg:socialem}
\end{algorithm}

Algorithm~\ref{alg:socialem} executes one EM iteration to find the next model parameters $\theta_{x+1}$. Notice that we do not execute the E-step separately. Because we only need to compute $\Pr(f,z|u, i)$ once for each user-item pair observed in $H$, we embed the E-step computation in the M-step so the posteriors are computed only as needed. Therefore, for each observed $\langle u,i\rangle$, the E-step is executed once in line \ref{socialem:estep}, and the M-step is executed  from line~\ref{socialem:mstep} to accumulate latent variable posteriors into the corresponding posterior sums (e.g. $\Pr^+(f|z)$ now takes the sum of all the posteriors with the same friend and topic ids). After all the observed user-item pairs are examined, M-step need to normalize posterior sums as next iteration's parameters in line \ref{socialem:norm}. These accumulation and normalization steps realize the M-step in Equation~(\ref{eq:prmstep}). The running time for this EM algorithm is $O(|H|\cdot |Z|\cdot|\overline{F(u)}|)$, where $|H|$ is the total number of observed user-item pairs, $|Z|$ is the latent topic size and $|\overline{F(u)}|$ is the average number of friends per user.

\subsection{Parallelized Map-Reduce Algorithms}
In this section, we show how we decompose the Algorithm~\ref{alg:socialem} for parallel processing. Notice that there are three computation components in one EM iterations: 1) E-step to compute posteriors $\Pr(f,z|u,i)$; 2) Accumulate posteriors to posterior sums; and 3) Normalization step to obtain the model parameters for next iteration. Among them, 
Step 1 and 3 could not be parallelized because Step 1 requires the knowledge about all the related model parameters $\theta_{x}$ and Step 3 requires the entire set of $\theta_{x+1}$ for parameter normalization. Therefore, based on the design principle of Map-Reduce algorithms, we execute Step 3 of previous iteration along with Step 1. As such, the non-parallelizable components are combined to avoid overhead of another round of Map-Reduce to achieve the same task.

\begin{algorithm}[htb]
\small
\KwIn{Partial Dataset: ${H_x} = \{\langle u, i\rangle\}$, Un-normalized model parameters: $\theta_x = \{\Pr(u|f),\Pr(f|z),\Pr(i|z)\}$}
\KwOut{Intermediate probabilities in key value pairs.}
Normalize $\Pr(z), \Pr(f|z), \Pr(i|z), \Pr(u|f)$\;\label{mapper:norm}
\For{$\langle u, i \rangle \in H_x$\label{mapper:mstep}}  {
 \For{$f \in F(u)$}  {
        \For{$z \in Z$} {
            Compute $\Pr(f,z|u,i)$\;
            $\Pr(f|u,i) \gets \Pr(f|u,i) + \Pr(f,z|u,i)$\;
        }
        Emit $\textrm{key: }f, \textrm{value: }\langle \Pr(f,z_0|u,i), \Pr(f,z_1|u,i) \cdots\rangle$\label{mapper:park}\;
        Emit $\textrm{key: }i, \textrm{value: }\langle \Pr(f,z_0|u,i), \Pr(f,z_1|u,i) \cdots\rangle$\;
    }
Emit $\textrm{key: }u, \textrm{value: }\langle \Pr(f_0|u,i), \Pr(f_1|u,i)  \cdots\rangle$\label{mapper:parkend}\;
}
\label{socialem:endmstep}
\caption{\small{Social Influence EM Mapper Algorithm}}
\label{alg:socialmapper}
\end{algorithm}

\begin{algorithm}[htb]
\small
\KwIn{Grouped intermediate probabilities.}
\KwOut{Un-normalized next parameters: $\theta_{x+1} = \{\Pr^+(u|f),\Pr^+(f|z),\Pr^+(i|z)\}$}
\For{ $\textrm{key}=K, \textrm{values}=\langle V_0, V_1, \cdots \rangle$ from input \label{reducer:iter}}  {
    $V \gets \sum_{x} V_x$\;
    Emit $\textrm{key:}K, \textrm{value:} V$\;
}
\caption{\small{Social Influence EM Reducer Algorithm}}
\label{alg:socialreducer}
\end{algorithm}

The algorithms for Mapper side and Reducer side are shown in Algorithm~\ref{alg:socialmapper} and Algorithm~\ref{alg:socialreducer}, respectively. At its start, each mapper normalize posterior sums from previous results to construct model parameters $\theta_x$ (see line ~\ref{mapper:norm}). Then the user-item pairs are processed in parallel at different mappers (line~\ref{mapper:mstep}) because each mapper now has the same global knowledge of $\theta_x$. And the accumulation step is done in reducer to find posterior sums of for next iteration. 

Since each mapper only processes a portion of the user-item history, one mapper does not have the entire knowledge of $\langle u, i\rangle \in H$, and thus cannot accumulate correct posterior sums (e.g. $\Pr^+(f|z)$). To address this problem, we move the accumulation step to reducers (Algorithm~\ref{alg:socialreducer}). In particular, we park posteriors for a specific user, item or friend id together and emit the packed value (line~\ref{mapper:park}-\ref{mapper:parkend}). To ensure these posteriors are correctly accumulated in reducer, a standard shuffle-and-sort is performed to all the emitted key-value pairs. In this way, the Map-Reduce framework ensures that \emph{all the emitted values for the same key} are grouped and processed together in the same reducer.

Thanks to the shuffle-and-sort step, the reducer algorithm (Algorithm~\ref{alg:socialreducer}) is very simple, which only takes a sum of all the grouped values and output the sums. Let us take the $\Pr^+(f|z)$ computation as an example to understand how the reducer performs its task as desired. When a mapper run algorithm~\ref{alg:socialmapper}, only a part of $\langle u,i \rangle$ pairs are processed and the posteriors with respect to a particular $f$ are emitted to the mapper outputs. Although emitted key-value pairs with the same key ($f$) may come from different mappers, but they are grouped by key ($f$) after keys are shuffled and sorted. Because now all the values (packed with posteriors) for the same key $f$ are grouped, the reducer can simply sum the posteriors to find the correct posterior sums. Similar steps can be done for all the other parameter computations, i.e., ($\Pr^+(i|z)$ and $\Pr^+(u|f)$). A reducer does not need to differentiate key types, because the accumulation steps are the same for user/item/friend ids. Recall that these reducer outputs are not the $\theta_{x+1}$ yet. What left is the normalization for each posterior sums to do in the next Map-Reduce iteration.

The above Map-Reduce EM algorithms addresses the scalability issue in learning model parameters. We find the Map-Reduce framework is quite suitable for expediting our EM algorithm because the posterior computation and accumulation (which incur significant cost) can be done in parallel.

\up
\section{Unified Generative Model}\label{sec:extend}

As mentioned earlier, we aim to developed a new generative model to unify the ideas of social influence, collaborative filtering and content-based methods for item recommendation. 
In this section, we present the unified generative model developed for our ultimate goal.

\begin{figure}[thl]
\up\up\up
\centering
\includegraphics[width=2.5in]{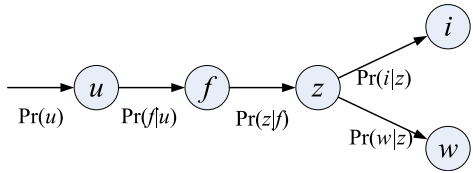}
\up
\caption{Unified generative model integrating social influence, collaborative filtering and content-based methods}\label{fig:cf-social-content}
\up\up\hup
\end{figure}

The unified generative model depicting a more general process of item selection is shown in Figure~\ref{fig:cf-social-content}. 
As shown, the early steps of the process is similar to the model introduced earlier (see Figure~\ref{fig:cf-social}). However, the selected topic now generates an item $i$ and a content description $w$. Therefore, a topic in this new model is not only associated with a distribution of items but also a distribution of item content (e.g., tag words).
Notice that here we assume items and contents are independently conditioned on the topics. As such, the similarity of item contents is taken into account in the recommendation process. As a result, the joint probability distribution over all factors is:  
\begin{equation}
\begin{split}
\Pr(u,f,z,i,w)  =  \Pr(u)\Pr(f|u)\Pr(z|f)\Pr(i|z)\Pr(w|z)
\end{split}
\end{equation}
where $w\in W$ and $W$ is the space of possible item contents. For example, $w$ could be a word of the content vocabulary or a tag of the tag space. Similar to Equation~(\ref{eqn:zifu}), the joint probability distribution can be rewritten as: 
\begin{equation}\label{eq:social}
\begin{split}
\Pr(u,f,z,i,w) & =  \Pr(z)\Pr(u|f)\Pr(f|z)\Pr(i|z)\Pr(w|z) \\
\end{split}
\end{equation}

Now the remaining issue is to learn the set of all the model parameters $\theta$. Different from what we discussed earlier in Section~\ref{sec:social}, the dataset used for learning now consists of three elements, including users, items and tags, i.e., $\langle u, i, w\rangle \in H$, where $u\in U$, $i\in I$, and $w\in W_i$ (i.e., $W_i$ denotes the tag/word set associated with item $i$). Note that an item may contain multiple tags/words.
For a history record of a user $u$ selecting an item $i$ where $W_i=\{w_1, w_2, \cdots\}$, we have $\langle u, i, w_k\rangle\in H$, $k = 1, 2, \cdots$.
Notice that $\theta$ now includes an extra parameter $P(w|z) (\forall z\in Z, w\in W)$ in addition to the other model parameters discussed earlier in our preliminary generative model (see Section~\ref{sec:social}).
The approach of learning model parameters is still to maximize the log-likelihood of $\mathcal{L}(\theta)$. 
However, the details are different.

In E-step, instead of computing the expectation of the log-likelihood for individual latent variables (e.g., $z$ or $f$ individually), we propose to compute the expectation of the log-likelihood for the joint latent latent variables (i.e., $z$ and $f$ together). More specifically, we calculate \up
\begin{equation}\label{eq:estep}
\begin{split}
&\Pr(z,f|u,i,w) \\= &\frac{\Pr(z) \Pr(f|z)\Pr(u|f) \Pr(i|z)\Pr(w|z)}{\sum_{z\in Z}\sum_{f\in F(u)} \Pr(z) \Pr(f|z)\Pr(u|f) \Pr(i|z)\Pr(w|z)}
\end{split}
\up
\end{equation}

In M-step, model parameters are computed to maximize the expected log-likelihood found on the E-step as below.

\up
\begin{equation}
\begin{split}
\Pr^+(i|z) & \propto \sum_{\langle u',i, w'\rangle \in H}\sum_{f' \in F(u')} \Pr(z,f'|u',i,w')\\
\Pr^+(w|z) & \propto \sum_{\langle u',i', w\rangle \in H}\sum_{f' \in F(u')} \Pr(z,f'|u',i',w)\\
\Pr^+(u|f) & \propto \sum_{\langle u,i',w'\rangle \in H \wedge f\in F(u)}\sum_{z' \in Z} \Pr(z',f|u,i',w')\\
\Pr^+(f|z) & \propto \sum_{\langle u',i',w'\rangle \in H \wedge f\in F(u')} \Pr(z,f|u',i', w') \\
\Pr^+(z) & \propto \sum_{\langle u',i', w'\rangle \in H}\sum_{f' \in F(u')} \Pr(z,f'|u',i', w')
\end{split}
\up
\end{equation}\label{eq:pr-mstep}
The parameters maximization method is similar to Equation~(\ref{eq:prmstep}). Note that the summed latent variable posterior is different and that we have an additional set of parameters in $\Pr^+(w|z)$.

After the model is learned, items can be ranked for a given user based on $\Pr(i|u)$, which can be approximated by Equation~(\ref{eq:cf_rec}), in which 
\begin{equation*}
\begin{split}
\Pr(u,i) = \sum_{z\in Z}\sum_{f\in F(u)}\sum_{w\in W_i}\Pr(z)\Pr(i|z)\Pr(f|z)\Pr(u|f)\Pr(w|z).
\end{split}\up
\end{equation*}
Note that we are only interested in the tags/words associated with the given item when we calculate the user-item joint probability. Item $i$ with high $\Pr(i|u)$ that user $u$ has not yet accessed is a good candidate for recommendation.

\up

\section{Group Recommendation}\label{sec:group}
Given a group of people $G$, group recommendation aims to find items that are welcomed by the whole group instead of individual group members. This recommendation service has a very large application base, e.g., coordinating a group of people to find quality activities/venues/restaurants/movies, etc. Although the generative models we proposed earlier are targeting on item recommendation for an individual user, the social influence parameter learned in our models is very useful for group recommendation. In this section, we first introduce the state-of-the-art approaches for group recommendation, namely \emph{aggregation-based  group recommendation} and then discuss how to apply the quantified social influence obtained from our models to develop a new algorithm, called \emph{social influence based  group recommendation}, for group recommendation.

\subsection{Aggregation-based Recommendation}
For group recommendation, one popular approach is the \emph{ranking aggregation} method which finds a ``consensus'' ranking/score for each item for the whole group. Given individual ranking/score for each member, some aggregation methods are employed to obtain a group ranking/score from individual ranking/scores. In this paper, we review two popular aggregation strategies: \emph{average} and \emph{least misery} recent proposed in~\cite{Masthoff04UMUAI}.

\textbf{Average} - With the average aggregation strategy, an item $i$'s group score is defined as an \emph{average} of the scores from individual group members. By using the item access probability estimation $\Pr(i|u)$ as the score, the group score for an item $i$ to group $G$ is calculated as
\begin{equation}
S_\textrm{average}( G, i)=\frac{\sum_{u\in G}\Pr(i|u)}{|G|}
\end{equation}
Accordingly, the recommendation ranking can be computed by sorting the group scores in descending order.

\textbf{Least Misery} - With the least misery aggregation strategy, the group score for item $i$ to a group $G$ is equal to the \emph{smallest} predicted rating for item $i$ in the group, specifically
\begin{equation}
S_\textrm{misery}( G, i) = \min_{u\in G}\{\Pr(i|u)\}
\end{equation}
Following this strategy, whether an item is acceptable to the group depends on the least satisfied member. Basically, the item least disliked by each individual member shall has the highest group score for recommendation.

These two aggregation-based group recommendation approaches captures a group consensus of item ranking by considering all the decisions made by users to be independent and equally important. However, in a group activity, people interact with each other and thus influence each other. We aim to address this observation in our social influence based group recommendation algorithm.

\subsection{Social Influence based Recommendation}

Note that we restrict the recommendation to a group where every group member has at least one friend in the group.
Within such a group, friends may influence each other so there may exist a group consensus. While our generative models aim to capture the process where a given user $u$ (influenced by a friend $f$) selects an item $i$, we can also see the process as a group activity, 
i.e., $u$ is influenced by $f$ to jointly select item $i$. Intuitively, our models can be used directly to support group recommendation for ``two-member'' groups.

Let $G_2=\{u_1,u_2\}$ denote a ``two-member" group. To select an item for the group, user $u_1$ could influence user $u_2$ and vice versa. Therefore, we define the score for recommending an item $i$ to the group $G_2$ as
\begin{equation}
S_\textrm{influence}(G_2,i) = Pr(u_1,u_2,i) + Pr(u_2,u_1,i)
\end{equation}
where
\begin{equation} \label{eqn:group-social-2}
\begin{split}
\Pr(u,f,i) &= \sum_{z\in Z} \Pr(u,f,z,i)\\
& = \sum_{z\in Z} \Pr(z)\Pr(u|f)\Pr(f|z)\Pr(i|z)
\end{split}
\end{equation}
can be easily obtained from the model parameters of our generative models.

\begin{figure}[thl]
\centering
\includegraphics[width=2.8in]{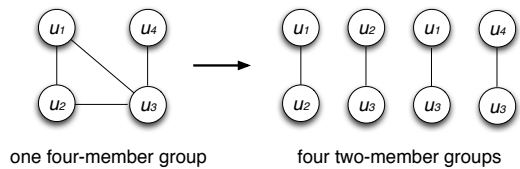}
\caption{Decompose an arbitrary group into a set of two-member groups. Edges between nodes denote online friendship.}\label{fig:group}
\end{figure}

The ideas described above can be generalized for groups with more than two members. A group of more than two members can be decomposed into a set of two-member groups based on the friendship of members (see the example in Figure~\ref{fig:group} for illustration). To make a group recommendation, we assume the social influence only takes action between friends. Intuitively, if most pairs of friends in the group prefer a particular item, it would be a good candidate for recommendation to the group. Let $G$ denote a group with arbitrary cardinality. The score for recommending an item $i$ to $G$ is defined as the sum of $S_\textrm{influence}(G_2,i)$ score over all possible friend pairs in the group.
Formally,

\begin{equation}
\begin{split}
S_\textrm{influence}( G, i) & = \sum_{\forall \langle u,f\rangle\in G\times G, u\neq f, f\in F(u)}S_\textrm{influence}(\{u,f\},i)
\end{split}
\end{equation}
The ranking of items for group recommendation is based on the sorted group scores of items as defined above. Thus, the decision of selecting an item for the group naturally incorporates the social influence among members of the group. We find superior performance of our social-influence strategy over the two aggregation strategies (to be shown in our evaluation).

\up
\section{Performance Evaluation}\label{sec:perform}
In this section, we validate our proposed probabilistic generative models using two real datasets, one from \emph{last.fm} and the other from \emph{whrrl.com}. We develop web crawlers to collect theses two datasets, which include user-item accessing history, users' friendship network and tags associated with each item.
Besides, we collect group check-in history data from whrrl.com to validate our group recommendation approach. In our evaluation, we adopt the user-based collaborative filtering approach (denoted as CF) as a baseline and propose to study the effectiveness of different factors (i.e., social influence, collaborative filtering and item content) included in our unified generative model. The different configurations of factors included on our evaluation are: 1) CF factor (CF-PGM) (see Figure~\ref{fig:cf}), 2) combination of CF and social influence factors (CF+SI-PGM) (see Figure~\ref{fig:cf-social}), 3) combination of CF and item content factors (CF+IC-PGM) (this has been discussed in \cite{PopesculUPL01UAI}), and 4) combination of CF, social influence and item content factors (CF+SI+IC-PGM) (i.e., our unified model; see Figure~\ref{fig:cf-social-content}). In this evaluation, we conduct a comprehensive set of experiments for item recommendation (to a single user) and group recommendation.

\subsection{Dataset Description}
\begin{figure}[htb]
  \subfigure[User Items] {\includegraphics[width=1.06 in]{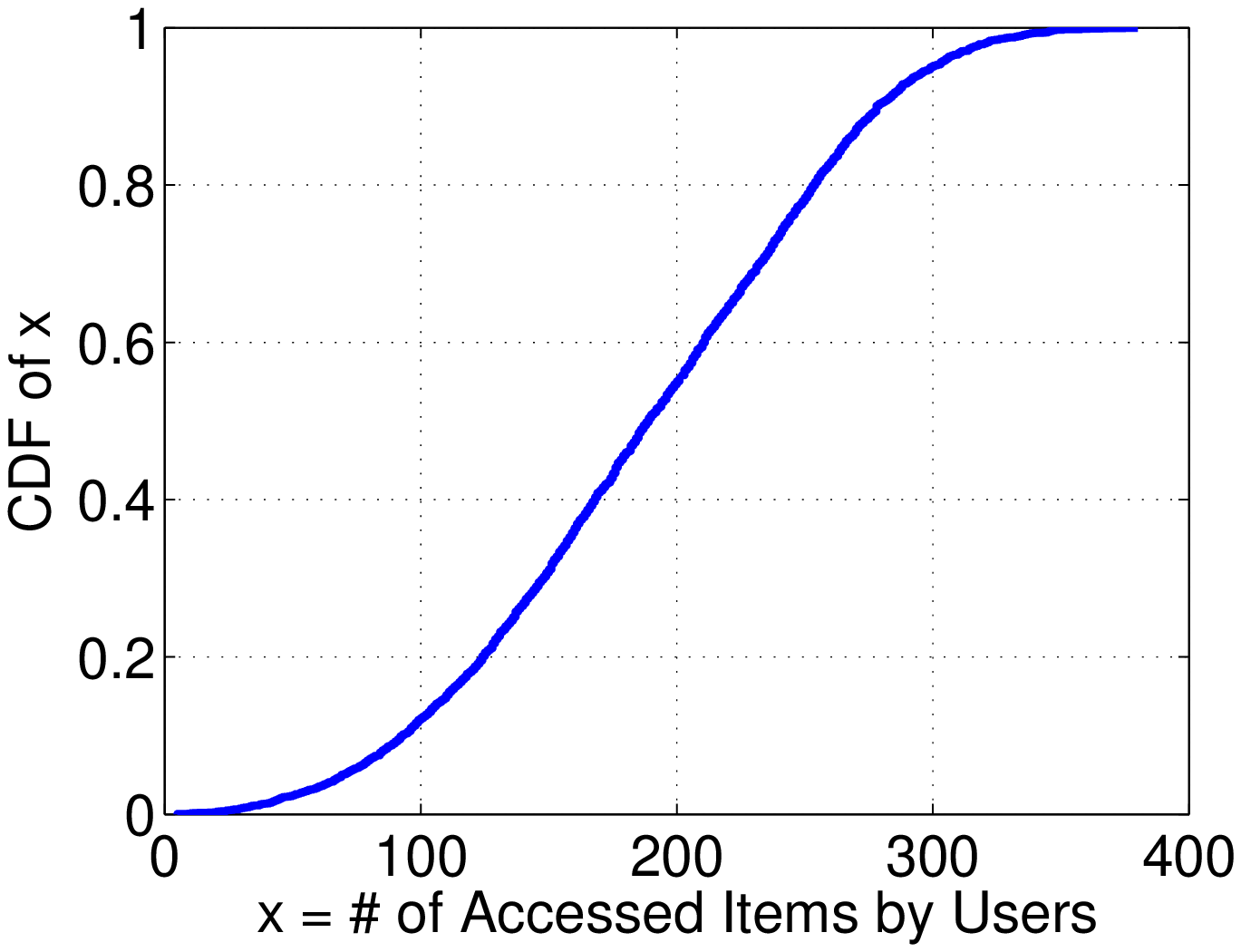}}
  \subfigure[User Friends] {\includegraphics[width=1.06 in]{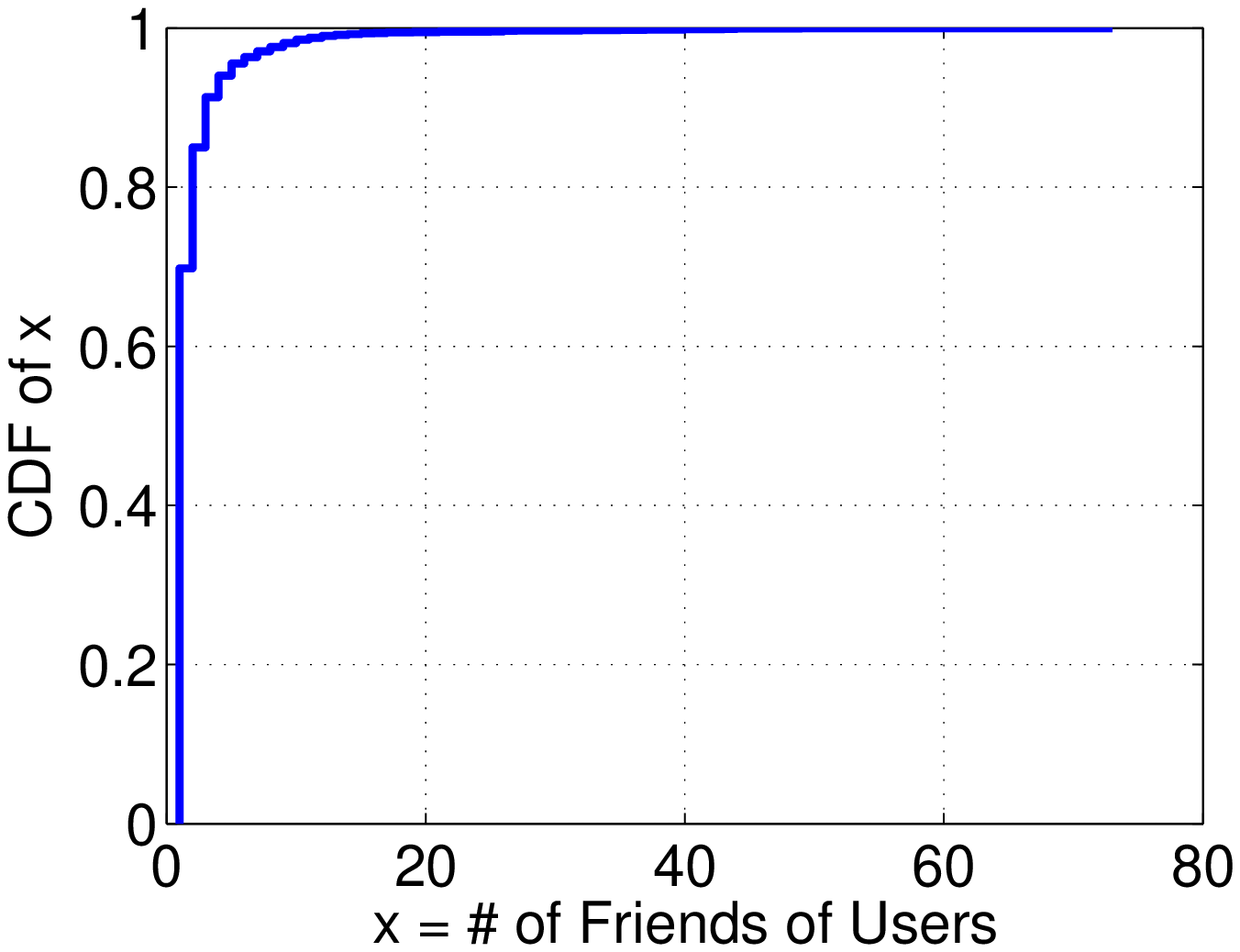}}
  \subfigure[Item Tags] {\includegraphics[width=1.06 in]{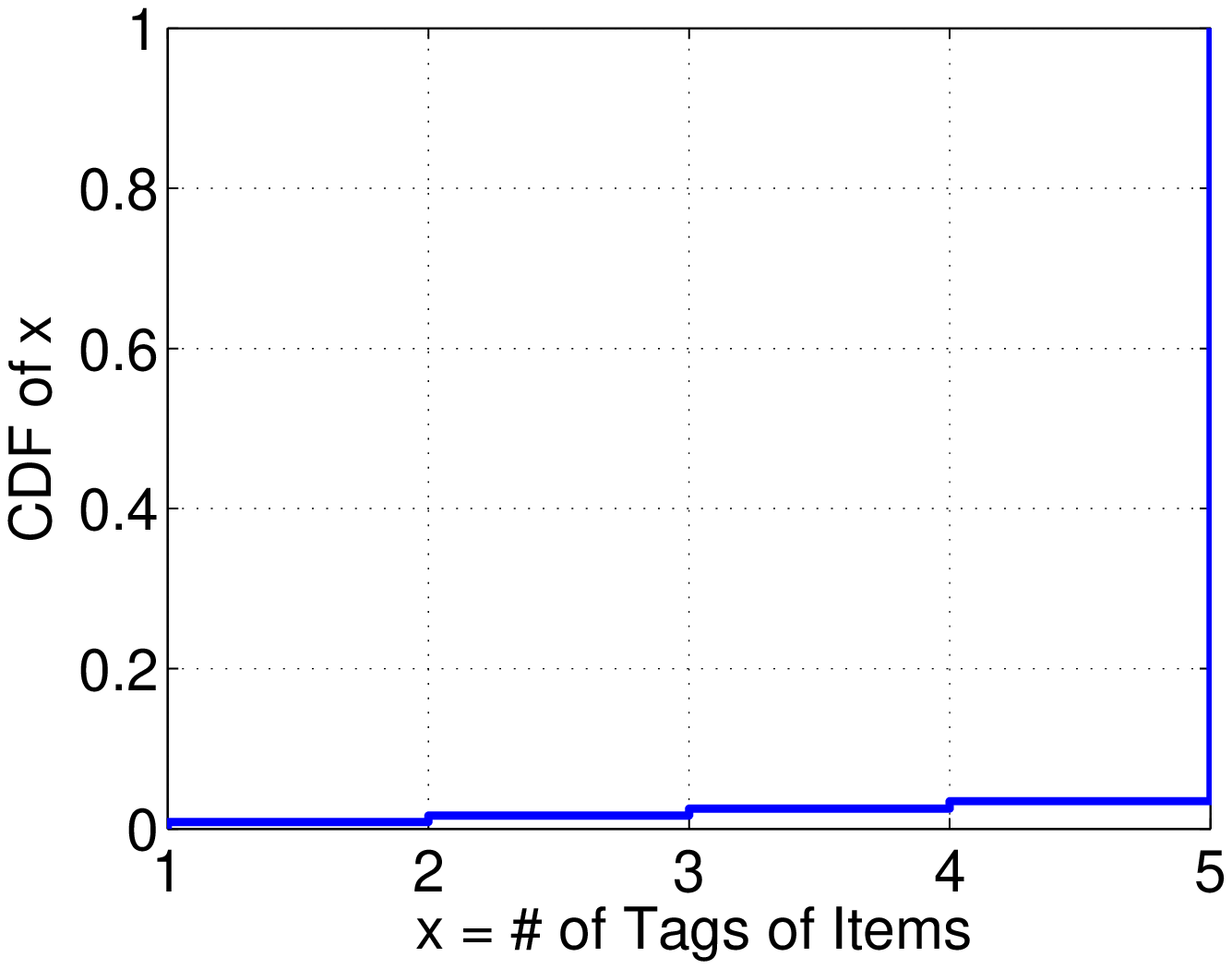}}
  \caption{Cumulative Distributions (last.fm)}\label{fig:lastfm}
\end{figure}

\begin{figure}[htb]
  \subfigure[User Items] {\includegraphics[width=1.06 in]{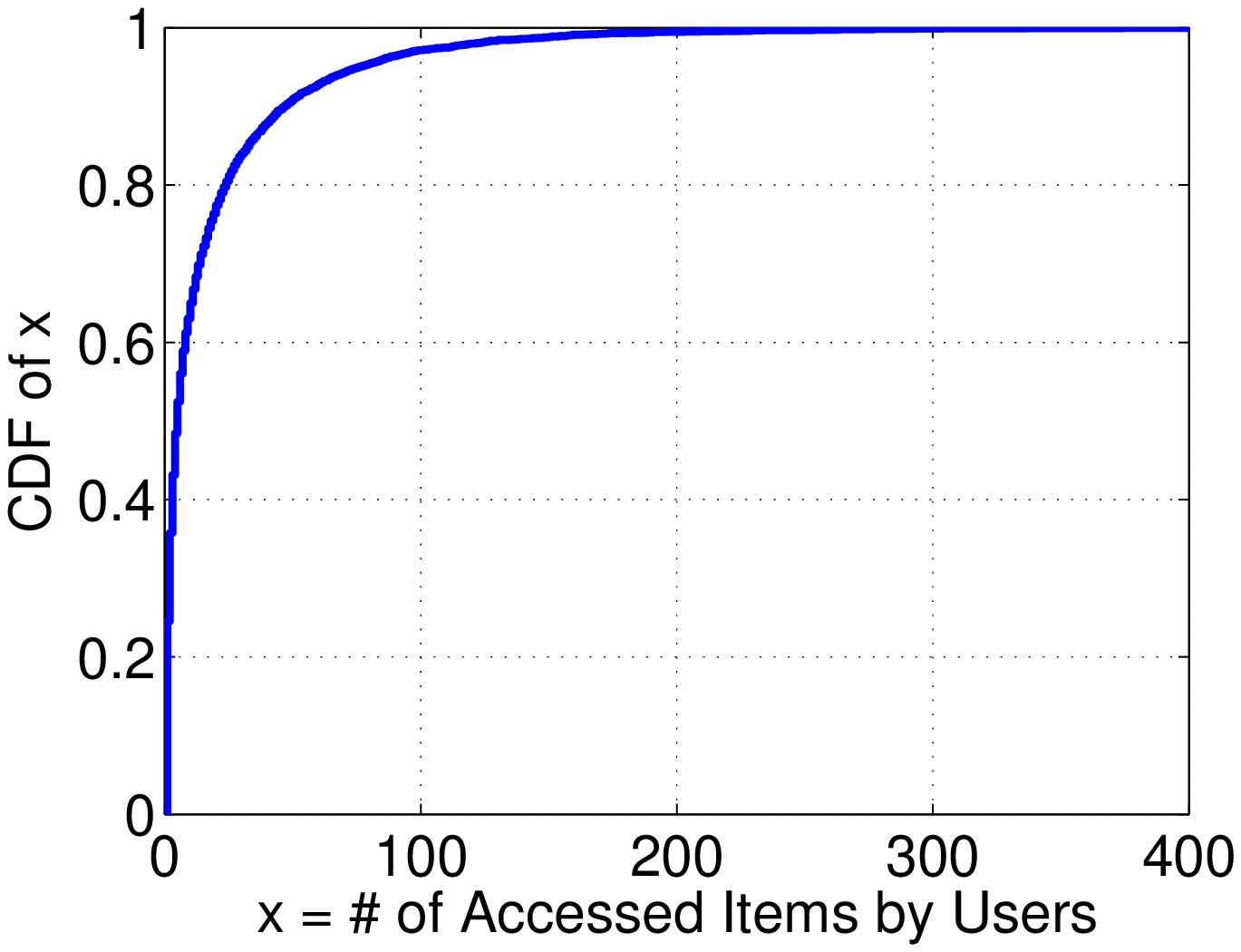}}
  \subfigure[User Friends] {\includegraphics[width=1.06 in]{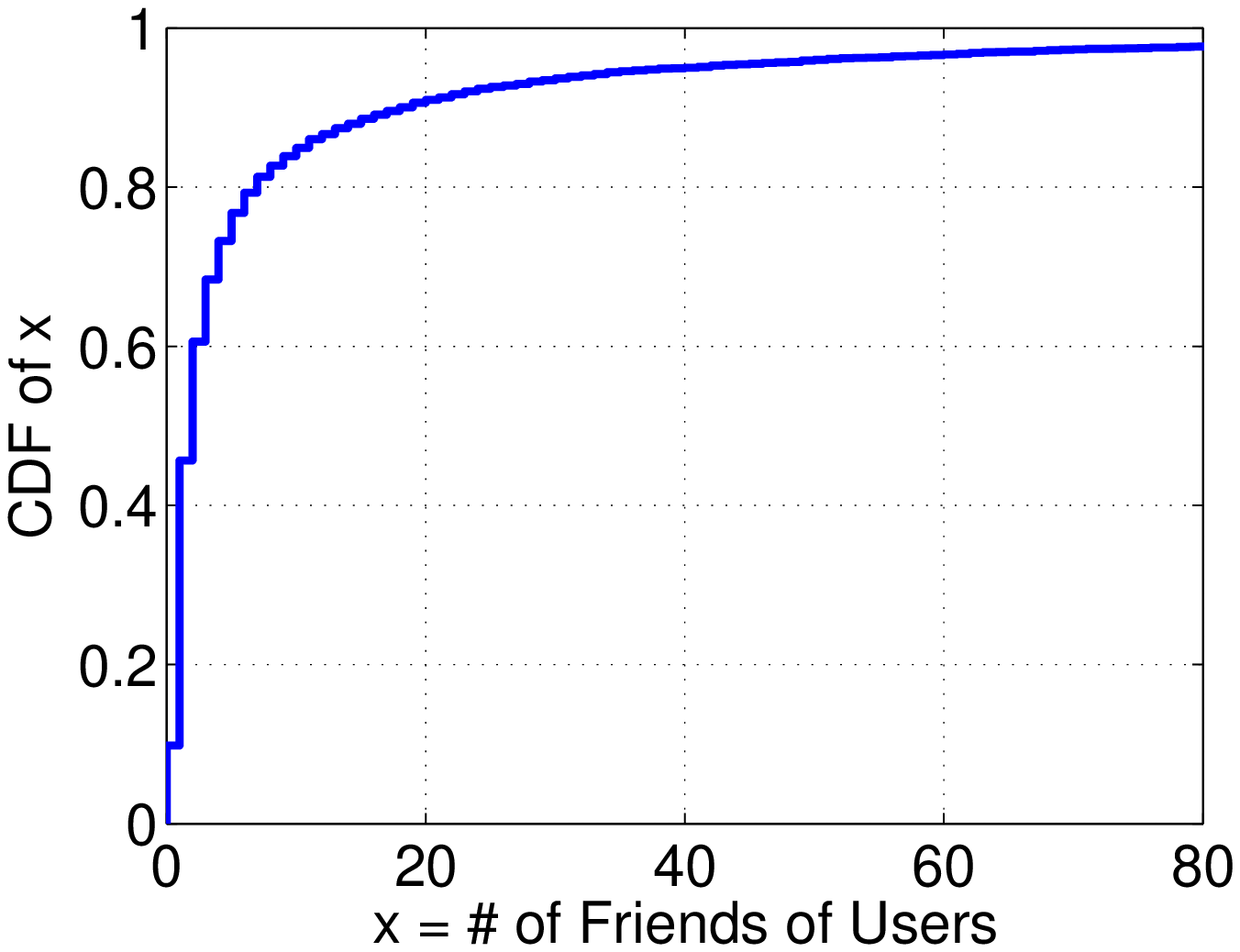}}
  \subfigure[Item Tags] {\includegraphics[width=1.06 in]{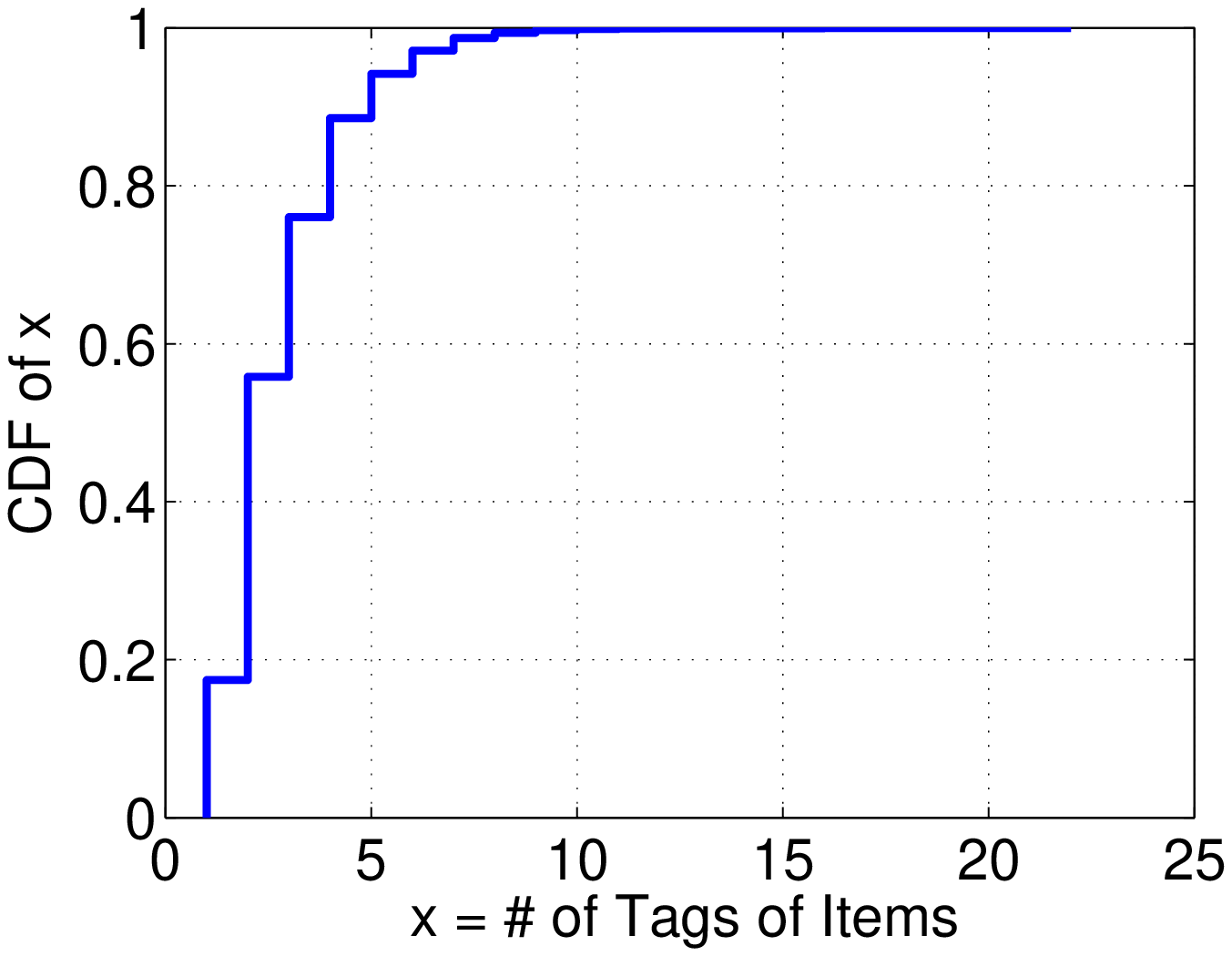}}
  \caption{Cumulative Distributions (whrrl.com)}\label{fig:whrrl}
\end{figure}

Here we first provide information about the datasets, i.e., \emph{last.fm} and \emph{whrrl.com}, used in our experiments. Last.fm is an on-line music radio web service and whrrl.com is a location-based social network web service. The last.fm dataset contains music access history of $3,143$ users over $23,467$ unique songs; while whrrl.com dataset includes the check-in history of $7,145$ users to $74,217$ unique places. It is worth noting that the whrrl.com dataset includes 17,587 \emph{group} check-in records which are very valuable for evaluating the group recommendation approaches. Additionally, both datasets have their user social networks available.
The basic statistics of these two datasets are summarized in Table~\ref{tbl:datasets}. Cumulative distributions with respect to the number of items accessed by users (User Items), the number of friends of users (User Friends), and the number of tags associated with items (Item Tags) are shown in Figure~\ref{fig:lastfm} and Figure~\ref{fig:whrrl} for last.fm and whrrl.com, respectively.

\begin{table}
\begin{tabular}{|r|c|c|}
  \hline
  & last.fm & whrrl.com \\ \hline\hline
  Number of Users & 3,143 &  7,145\\ \hline
  Number of Items & 23,467 &  74,217\\ \hline
  User-Item Matrix Density & $8.02\times10^{-3}$ &  $2.3\times10^{-4}$ \\ \hline
  Average Friends per User & 1.91 & 9.08 \\ \hline
  Average Tags per Item & 4.92 & 2.73 \\ \hline
  Average Group Size & N/A & 2.93 \\ \hline
  \hline
\end{tabular}
\caption{Datasets Statistics}\label{tbl:datasets}
\end{table}

\subsection{Parameter Initialization and Training}
After the datasets are prepared, we are able to apply EM algorithms to infer model parameters. However, for all EM methods, model parameters need to be initialized and the iteration termination condition needs to be specified. After experimenting with different model parameter initialization methods, we decided to use Latent Dirichlet Allocation (LDA)~\cite{lda} to initialize the model parameters. Although LDA has been mainly used for clustering documents, it has similar parameters as the CF-PGM in Figure~\ref{fig:cf}. To obtain initial parameters from LDA, we treat each user as a ``document'' in the LDA model and transform items accessed by the user as ``words'' in the document. After the LDA model converges, we discard the document clustering but keep $\Pr(u|z)$ and $\Pr(z|i)$ as the model initialization values for EM algorithms. For the social influence parameters $\Pr(u|f)$ required in CF+SI-PGM, we use the normalized Jaccard similarity as the initial values (i.e., $Pr(u|f) \propto \textrm{JaccardSim}(I(u),I(f))$, where $I(u)$ and $I(f)$ denote the accessed items for $u$ and $f$, respectively). This initialization ensures friends having more commonly accessed items to have a higher social influence initially. Note that in our models, $u$ is treated as a special friend of himself. Since $\textrm{JaccardSim}(V(f),V(f))\equiv1$, a user's self influence is always larger than any social influence from his friends at the beginning. To terminate the EM algorithms, we use log-likelihood as model converge indicators and terminate the EM algorithms when an additional EM iteration cannot improve the training data's log-likelihood by $0.0001$ or when the maximum iteration threshold (empirically set with 50) is reached.

We implement both the single machine EM-algorithm and its Map-Reduce version and confirm that both implementations produce the same results with small datasets. For those more complicated models (i.e., CF+SI-PGM and \\ CF+SI+IC-PGM), we apply our Map-Reduce implementation on a cluster of 10 machines to infer the model parameters.

\subsection{Item Recommendation}
We use item recommendation as the primary test case to evaluate the performance of the probabilistic generative models under evaluation. We apply cross-validation method to find item recommendation's precisions and recalls. For both datasets, we mark off $30\%$ item assess history of each user for testing. In other words, the rest $70\%$ user-item pairs are used as training data to infer model parameters. Then after each model is learned, we use the model parameters to find $\forall i, \Pr(i|u)$ for all users. The items \emph{not in presence} in the training dataset are ranked based on their $(\Pr(i|u))$. In this way, we prevent our recommendation system from ``repeating'' a user's item access history. Therefore, all the recommendations for a user must be ``fresh'' items that have not been accessed by him in the training dataset. The precisions and recalls for top $n$ recommendations are used as the evaluation metrics, where $n = 5, 10, 20, 50$ (5 is the default value). Precision is calculated as the ratio of the number of recommendation hits to the recommendation size; and recall is calculated as the ratio of the number of recommendation hits to the size of user's validation item set. Then the average precisions and recalls of different users serve as the evaluation metrics. 
\begin{figure}[htb]
  \subfigure[Precision] {\includegraphics[width=1.6 in]{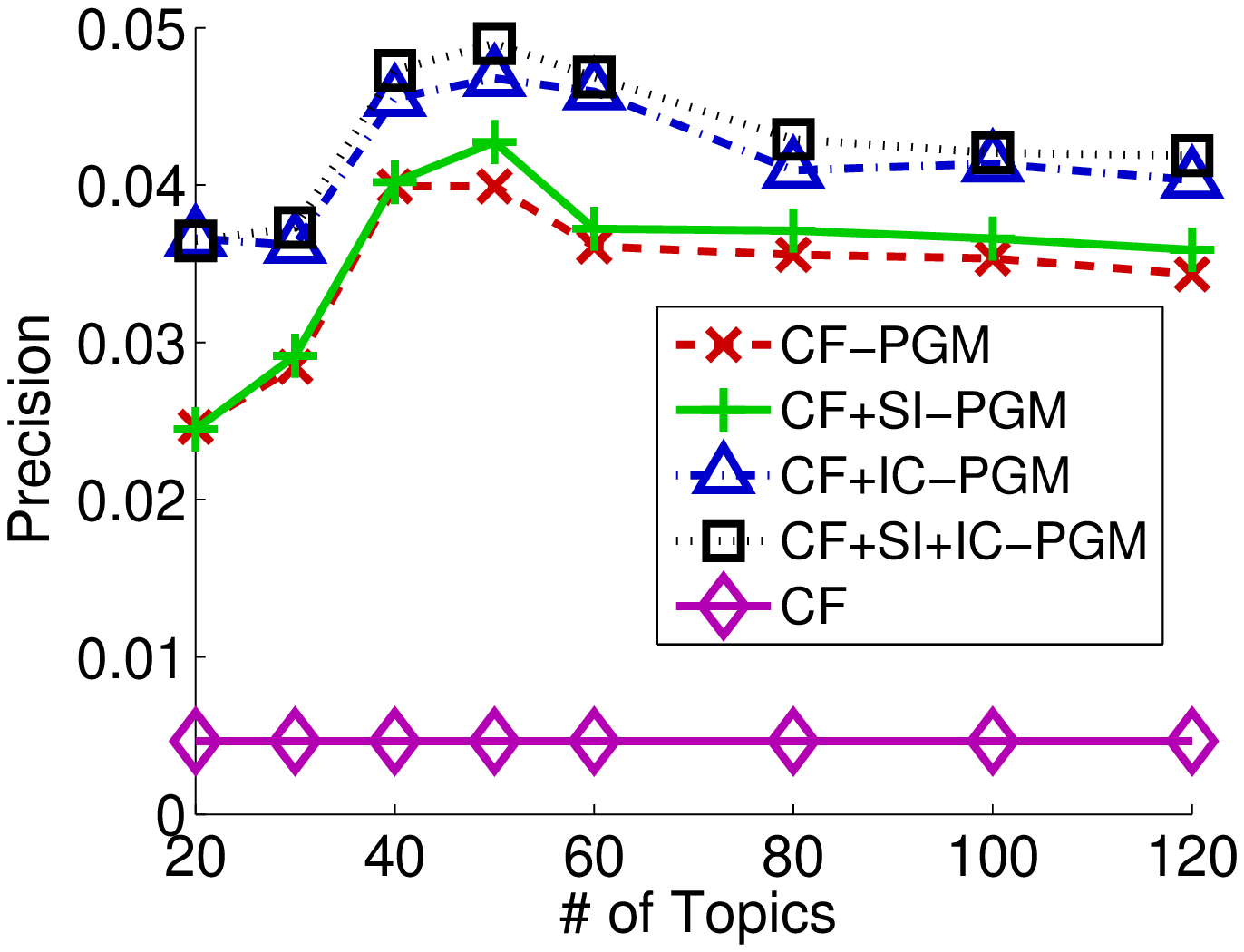}}\vspace{-0.1 in}
  \subfigure[Recall] {\includegraphics[width=1.6 in]{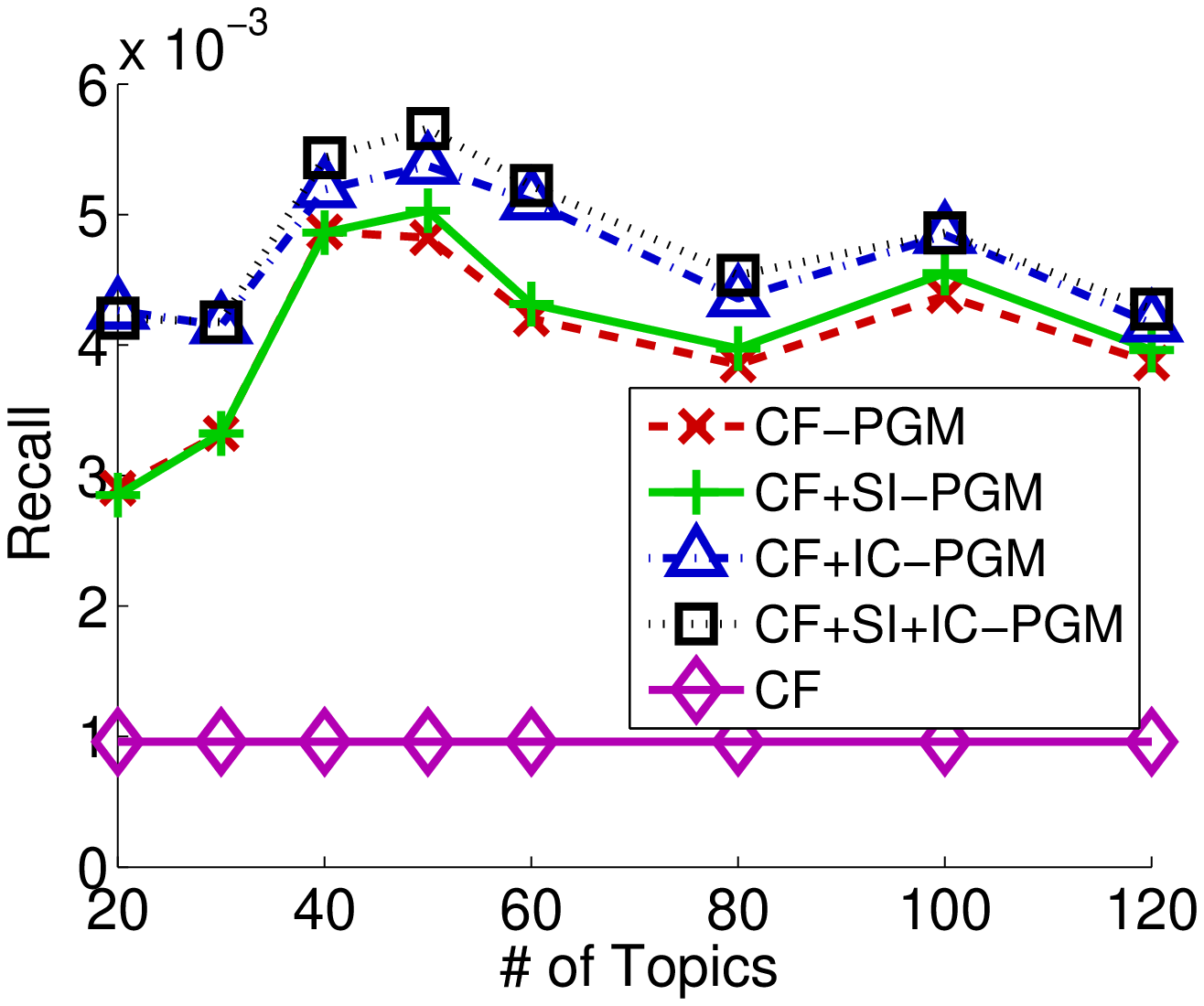}}
  \caption{Test Topic Sizes (last.fm)}\label{fig:lastfm_topic}
\end{figure}
\begin{figure}[htb]
  \subfigure[Precision] {\includegraphics[width=1.6 in]{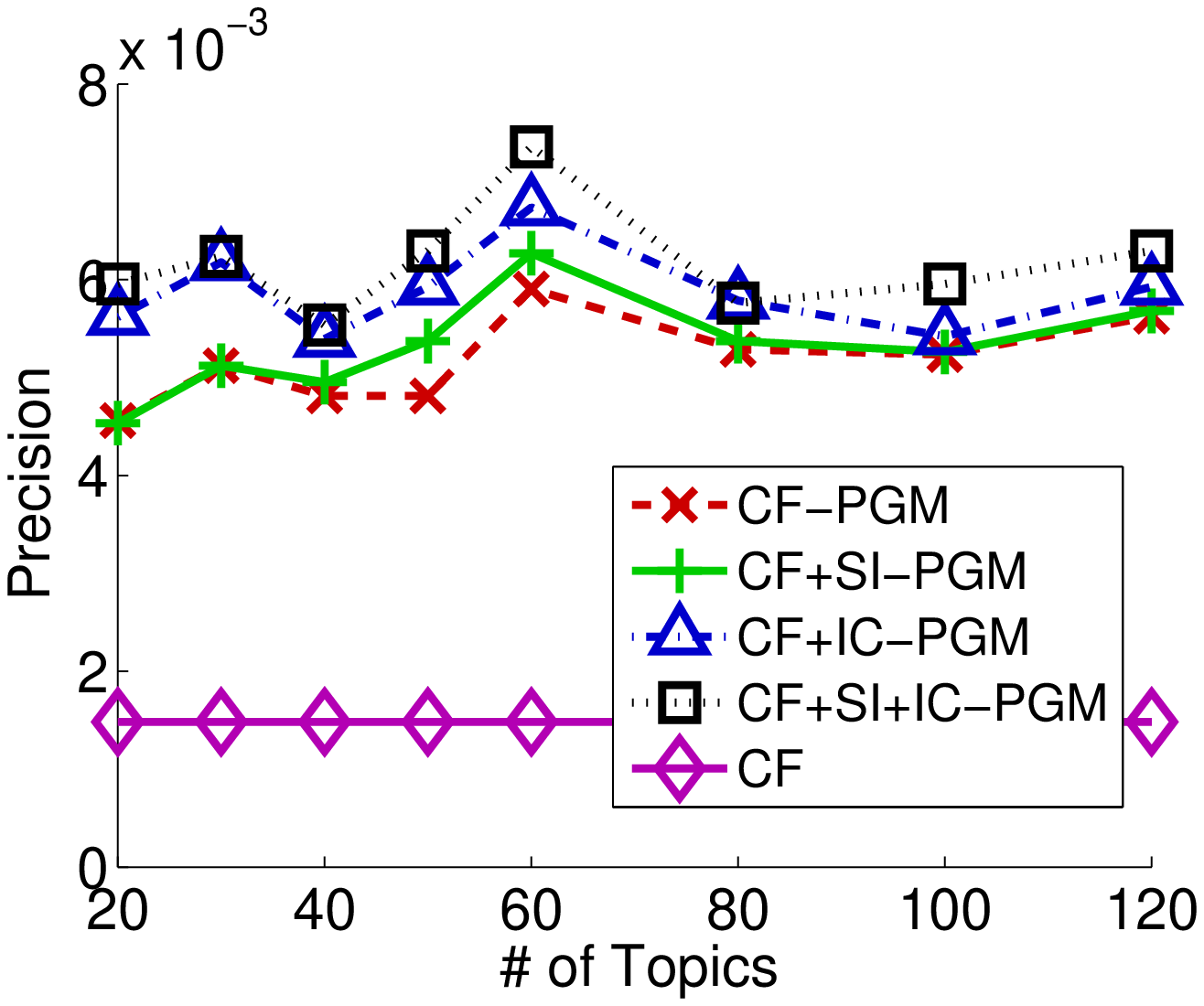}}\vspace{-0.1 in}
  \subfigure[Recall] {\includegraphics[width=1.6 in]{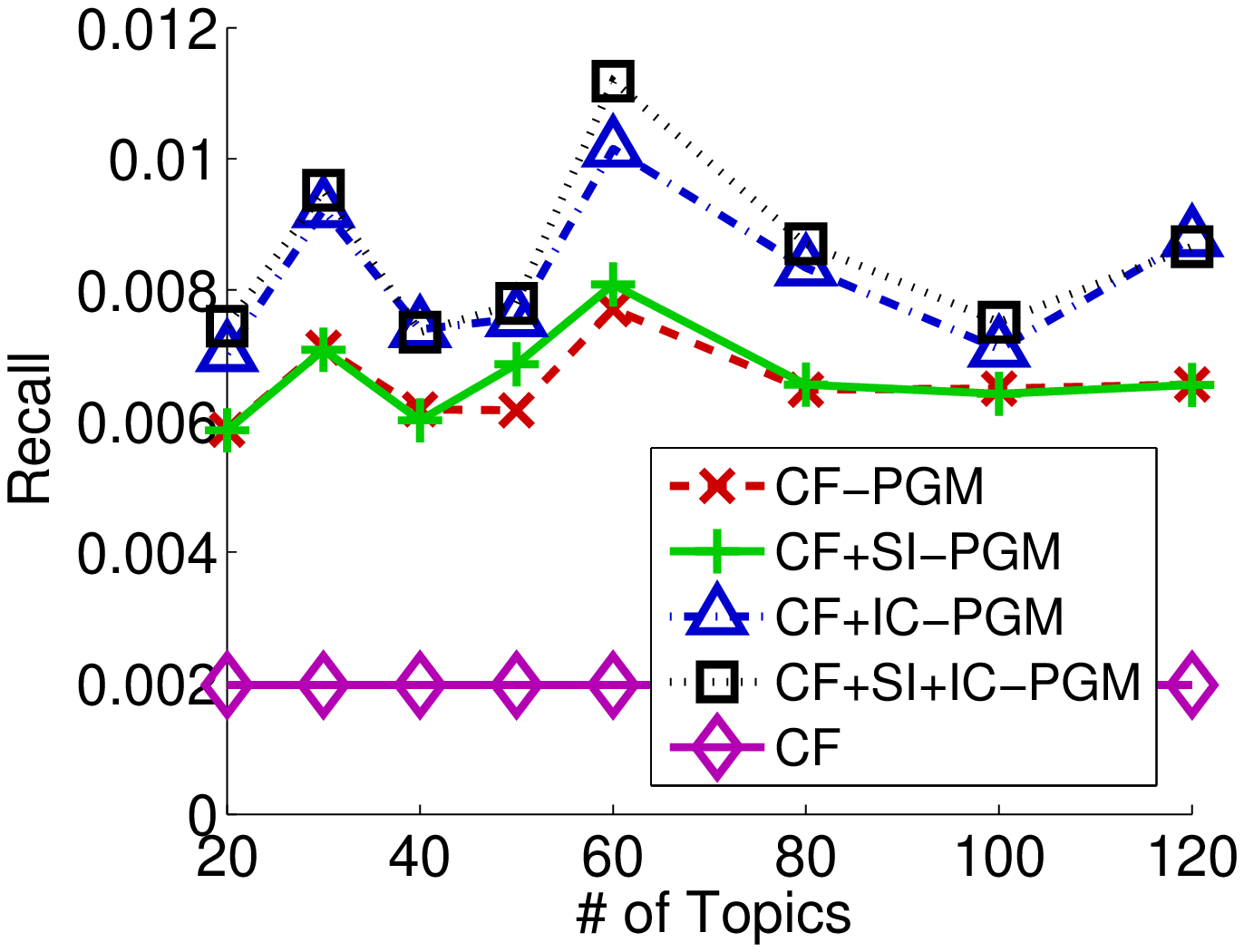}}
  \caption{Test Topic Sizes (whrrl.com)}\label{fig:whrrl_topic}
\end{figure}

In Figure~\ref{fig:lastfm_topic}~and Figure~\ref{fig:whrrl_topic}, the precision and recall of top 5 item recommendations for last.fm and whrrl.com under different latent topic sizes are presented. We find social influence indeed improves the recommendation performance, for both CF+SI-PGM against CF-PGM and CF+SI+IC-PGM against CF+IC-PGM. 
The result shows that the best recommendation performance is reached when the topic size is chosen around $60$. Therefore, we set the default value of the latent topic size to 60 for the remaining part of performance evaluation.

\begin{table}[htl]
  \centering
  \scriptsize
  \begin{tabular}{|c|r|c|c|c|c|}
    \hline
     & Methods & Top 5 & Top 10 & Top 20 & Top 50 \\ \hline \hline
    \multirow{5}{*}{\rotatebox{90}{\mbox{Precision}}}& CF-PGM & 0.0399 & 0.0372 & 0.0342 & 0.0296 \\ \cline{2-6}
    & CF+SI-PGM & 0.0427 & 0.0429 & 0.0383 & 0.0315 \\ \cline{2-6}
    & CF+IC-PGM & 0.0470 & 0.0542 & 0.0494 & 0.0388 \\ \cline{2-6}
    & CF+SI+IC-PGM & 0.0492 & 0.0566 & 0.0506 & 0.0398 \\ \cline{2-6}
    & CF & 0.0046 & 0.0066 & 0.0080 & 0.0085 \\ \hline \hline
    \multirow{5}{*}{\rotatebox{90}{\mbox{Recall}}}& CF-PGM & 0.0048 & 0.0085 & 0.0157 & 0.0329 \\  \cline{2-6}
    & CF+SI-PGM & 0.0050 & 0.0100 & 0.0187 & 0.0363 \\  \cline{2-6}
    & CF+IC-PGM & 0.0054 & 0.0102 & 0.0198 & 0.0385 \\  \cline{2-6}
    & CF+SI+IC-PGM & 0.0057 & 0.0117 & 0.0213 & 0.0411 \\  \cline{2-6}
    & CF & 0.0010 & 0.0024 & 0.0050 & 0.0122 \\  \cline{2-6}
    \hline
  \end{tabular}
  \caption{Performance on last.fm dataset}\label{tbl:lastfm_topn}

\vspace*{0.2in}

  \centering
  \scriptsize
  \begin{tabular}{|c|r|c|c|c|c|}
    \hline
     & Methods & Top 5 & Top 10 & Top 20 & Top 50 \\ \hline \hline
    \multirow{5}{*}{\rotatebox{90}{\mbox{Precision}}}& CF-PGM & 0.0048 & 0.0038 & 0.0035 & 0.0028 \\ \cline{2-6}
    & CF+SI-PGM & 0.0053 & 0.0041 & 0.0036 & 0.0028 \\ \cline{2-6}
    & CF+IC-PGM & 0.0059 & 0.0048 & 0.0040 & 0.0029 \\ \cline{2-6}
    & CF+SI+IC-PGM & 0.0062 & 0.0049 & 0.0041 & 0.0030 \\ \cline{2-6}
    & CF & 0.0015 & 0.0016 & 0.0015 & 0.0011 \\ \hline \hline
    \multirow{5}{*}{\rotatebox{90}{\mbox{Recall}}}& CF-PGM & 0.0062 & 0.0090 & 0.0141 & 0.0251 \\  \cline{2-6}
    & CF+SI-PGM & 0.0069 & 0.0100 & 0.0146 & 0.0254 \\  \cline{2-6}
    & CF+IC-PGM & 0.0076 & 0.0119 & 0.0157 & 0.0252 \\  \cline{2-6}
    & CF+SI+IC-PGM & 0.0081 & 0.0115 & 0.0154 & 0.0275 \\  \cline{2-6}
    & CF & 0.0020 & 0.0033 & 0.0051 & 0.0071\\  \cline{2-6}
    \hline
  \end{tabular}
  \caption{Performance on whrrl.com dataset}\label{tbl:whrrl_topn}
\end{table}

In Table~\ref{tbl:lastfm_topn}~and Table~\ref{tbl:whrrl_topn}, we compare the item recommendation performance of different algorithms. As shown in these two tables, all the probabilistic generative model approaches clearly outperform the conventional user-based collaborative filtering (CF).  Again, we find social influence factor indeed improves the recommendation performance, (for both CF+SI-PGM against CF-PGM and CF+SI+IC-PGM against CF+IC-PGM). Most importantly, the unified model we propose in this paper (which integrates collaborative filtering, social influence and item content) shows the best performance.

By comparing results from whrrl.com and last.fm datasets, we find that social influence is more important (in terms of item recommendation) in whrrl.com than last.fm.  One possible reason is that in our collected datasets, users in whrrl.com are more \emph{social} than users in last.fm, i.e., the average number of friends in whrrl.com is $9.08$ compared to last.fm's $1.91$. We also observe this phenomenon from the statistics shown in Figure~\ref{fig:lastfm}(b) and Figure~\ref{fig:whrrl}(b). In other words, it is more likely for users in whrrl.com to be influenced by their on-line friends than users in last.fm. 
Consequently, the recommendation performance benefit from social influence in last.fm is less significant than that in whrrl.com.

\subsection{Social Influence Study}
\begin{figure}[htb]
  \subfigure[Self Influence] {\includegraphics[width=1.6 in]{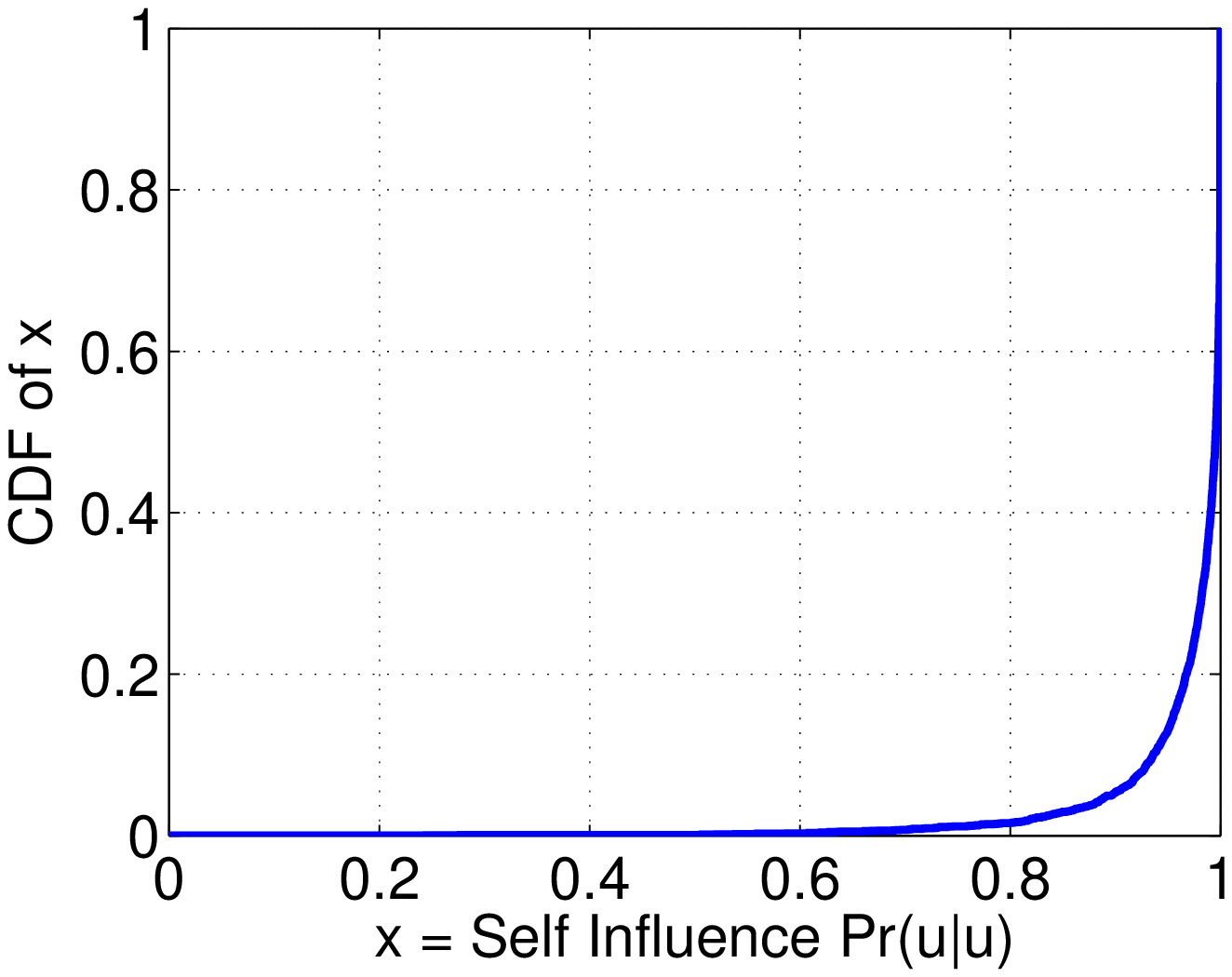}\label{fig:last_fm_self_influence}}\vspace{-0.1 in}
  \subfigure[Friend Influence] {\includegraphics[width=1.6 in]{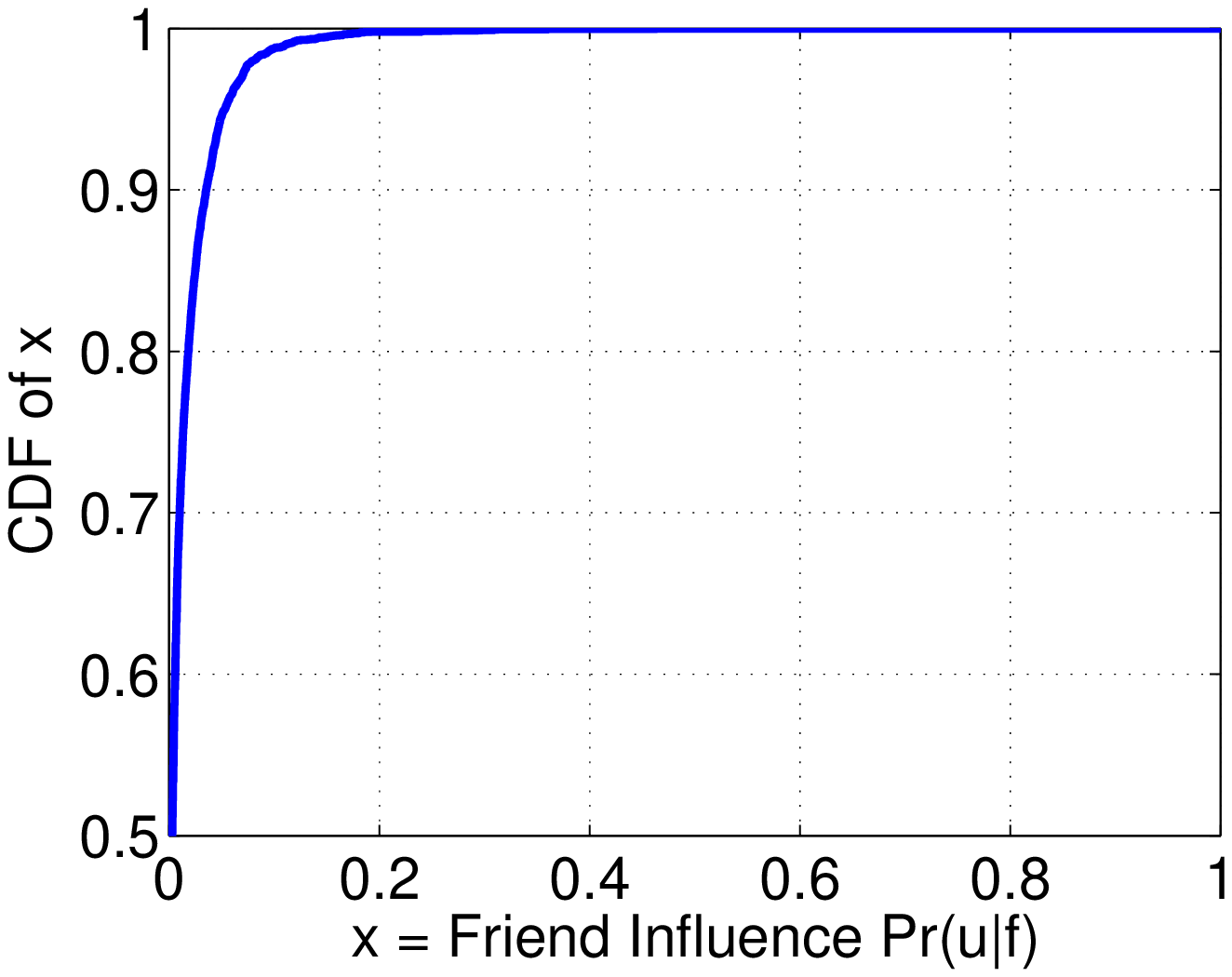}\label{fig:last_fm_friend_influence}}
  \caption{Social Influence Result (last.fm)}\label{fig:last_fm_influence}
\end{figure}

\begin{figure}[htb]
  \subfigure[Self Influence] {\includegraphics[width=1.6 in]{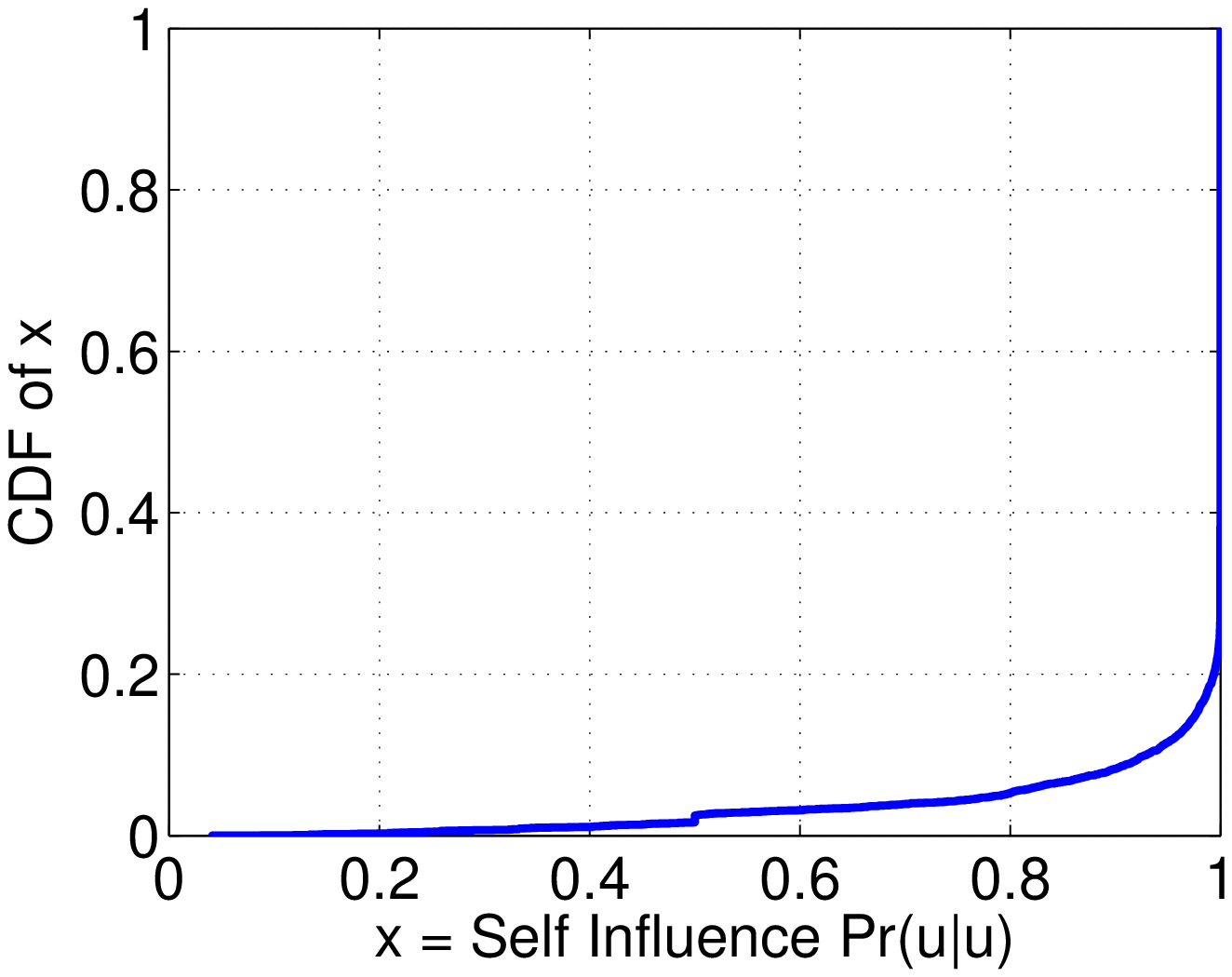}\label{fig:whrrl_self_influence}}\vspace{-0.1 in}
  \subfigure[Friend Influence] {\includegraphics[width=1.6 in]{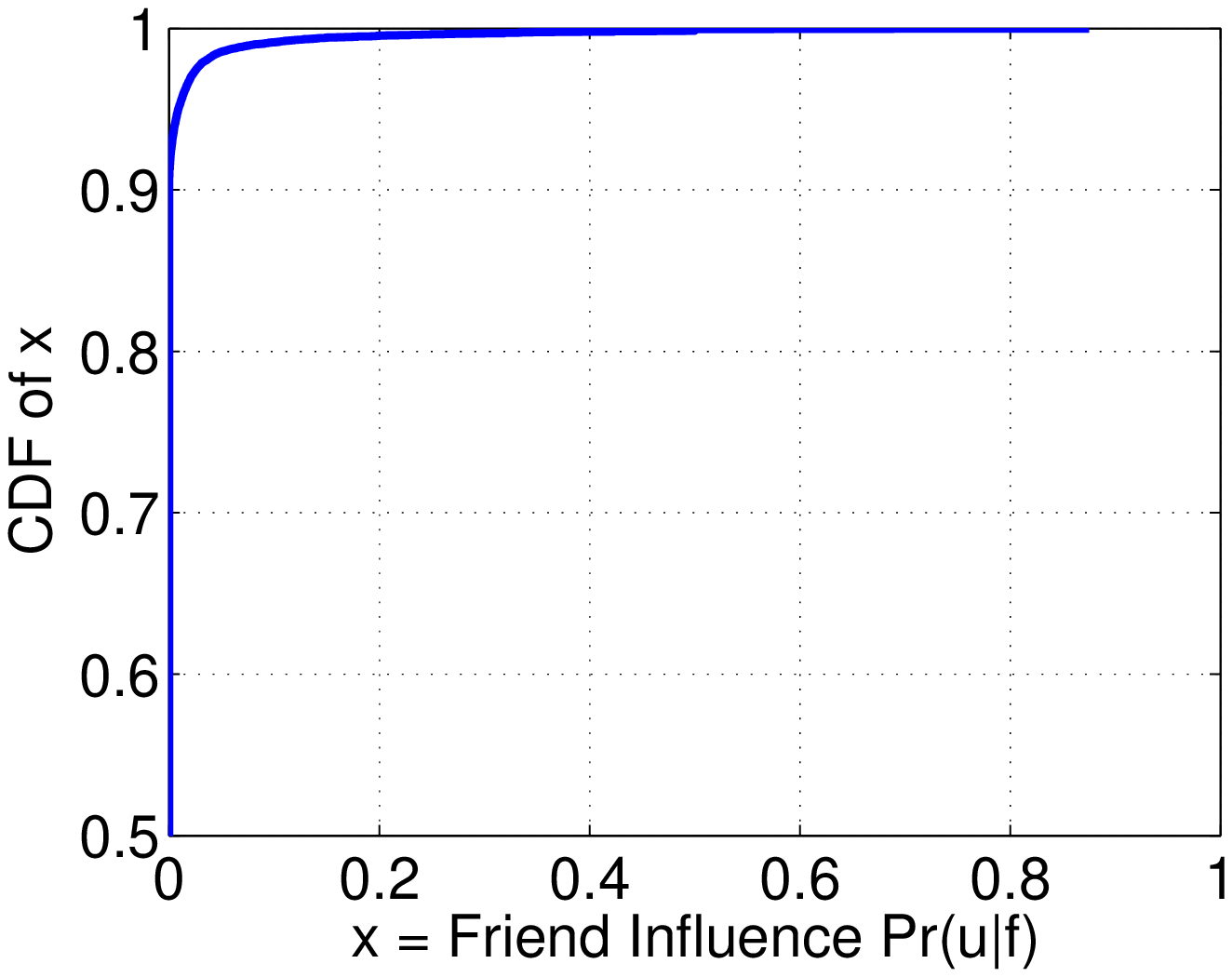}\label{fig:whrrl_friend_influence}}
  \caption{Social Influence Result (whrrl.com)}\label{fig:whrrl_influence}
\end{figure}

In this section, we aim to study the social influence between friends, where the social influence is learned through our proposed models. For simplicity, we focus on CF+SI-PGM. Instead of investigating how social influence improves the recommendation performance, here we are interested in how significant a particular user influence his friends. As different people have different personalities, we plot the distributions of social influence probabilities among friend pairs (that we learned through CF+SI-PGM) in Figure~\ref{fig:last_fm_influence} and Figure~\ref{fig:whrrl_influence}. Note that
we also consider the circumstance of self-influence and use $\Pr(u|u)$ to denote the probability.


Figure~\ref{fig:last_fm_influence}
shows the learned social influence presented in last.fm. In general, people's self-influence ($\Pr(u|u)$) in this dataset is significantly higher than the influence from their friends ($\Pr(u|f), f\neq u$) when choosing a music piece. Figure~\ref{fig:last_fm_self_influence} shows that more than $90\%$ users' self-influences are higher than $0.95$. Also, since each user may have several friends, each friend's social influence is thus quite small, e.g., $90\%$ friends' influence is smaller than $0.01$. This observation indicates that most music pieces consumed by the users in last.fm are selected in accordance with users' own preferences and tastes.

Figure~\ref{fig:whrrl_influence} demonstrates very different findings. As shown in Figure~\ref{fig:whrrl_self_influence}, the self-influence is still quite significant but much smaller than that in last.fm, i.e., $10\%$ users' self-influence is lower than $0.8$. The implication from this finding is that while people visit places mainly based on their own preferences, they would sometimes take friends' suggestions to visit places.
While users in whrrl.com have more friends than users in last.fm, (i.e., average $9.08$ friends compared to $1.91$), we find
a lot of friends are not influential
and that usually a small portion of friends takes the most part of social influence. In general, people's social influence in place check-in activities are much more significant than music consumption --- one explanation is that check-ins are inherently social activities and music consumption are usually for self-entertainment.

\subsection{Group Recommendation}
Finally, we report our findings on evaluation of group recommendation algorithms, including 
the social-influence based (SIG) algorithm introduced in Section~\ref{sec:group} along with two aggregation-based group recommendation strategies. To compare their performance, we use the $17,587$ group check-in records in whrrl.com. In our experiment, we consider a group check-in record (i.e., the ground truth) at a time and take the average of tested records. Notice that a record indicates a group of people visits a place. An effective group recommendation algorithm should have this place ranked high among all the places returned. Therefore, we propose a metric called \emph{relative ranking} to evaluate the performance of these group recommendation algorithms. Suppose that a given group recommendation algorithm returns a ranked list of $m$ items (i.e., places in this experiment). If the actual visited place is ranked in the $l$-th position of the returned list, the relative ranking is calculated as $\frac{l}{m}$.  For example, if an actual visited place is ranked 10th among a total of 100 items returned by a group recommendation algorithm, the relative ranking is $10/100=0.1$).
\begin{figure}[htb]
  \subfigure[Group Size] {\includegraphics[width=1.6 in]{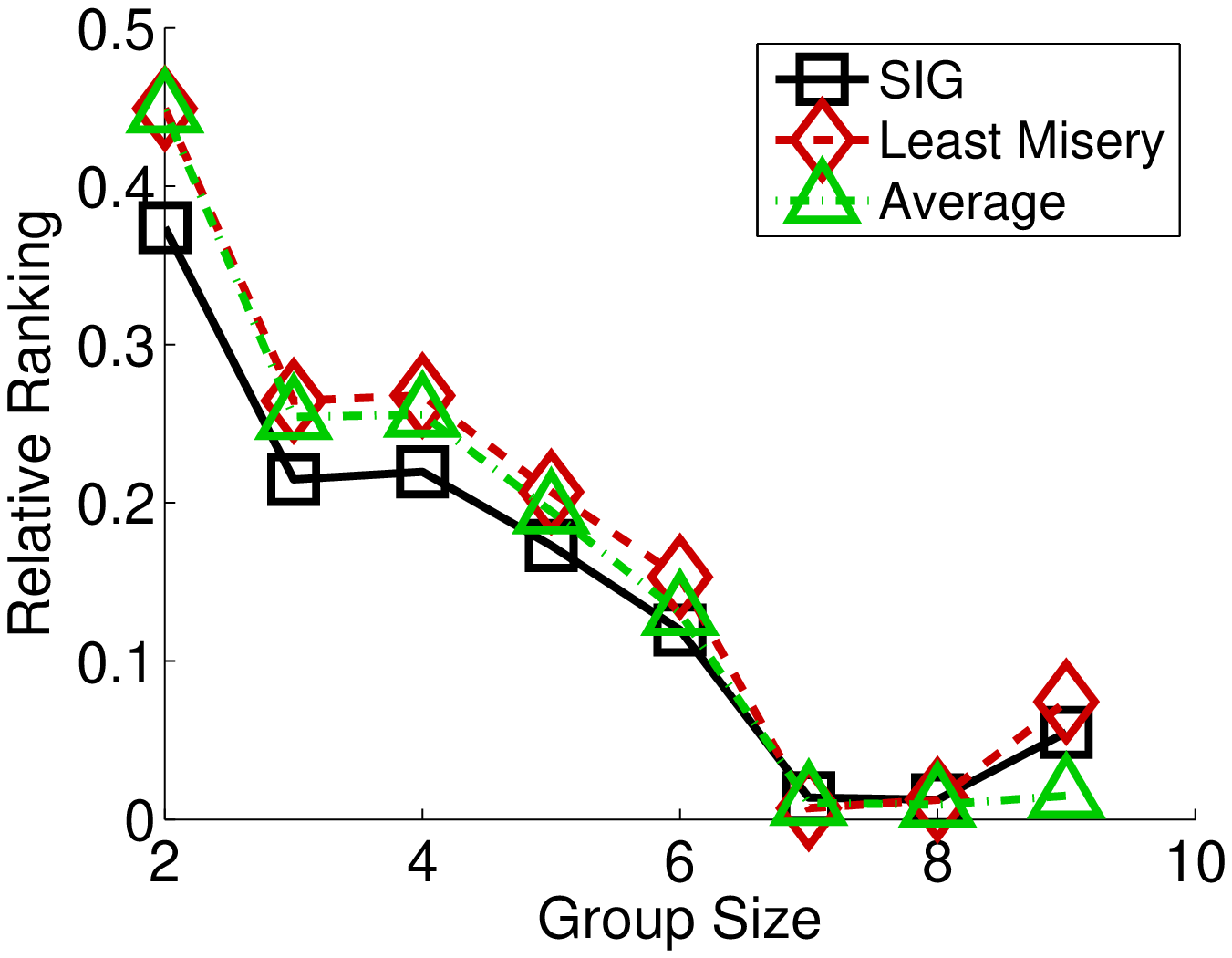}\label{fig:whrrl_group_size}}\vspace{-0.1 in}
  \subfigure[Topic Size] {\includegraphics[width=1.6 in]{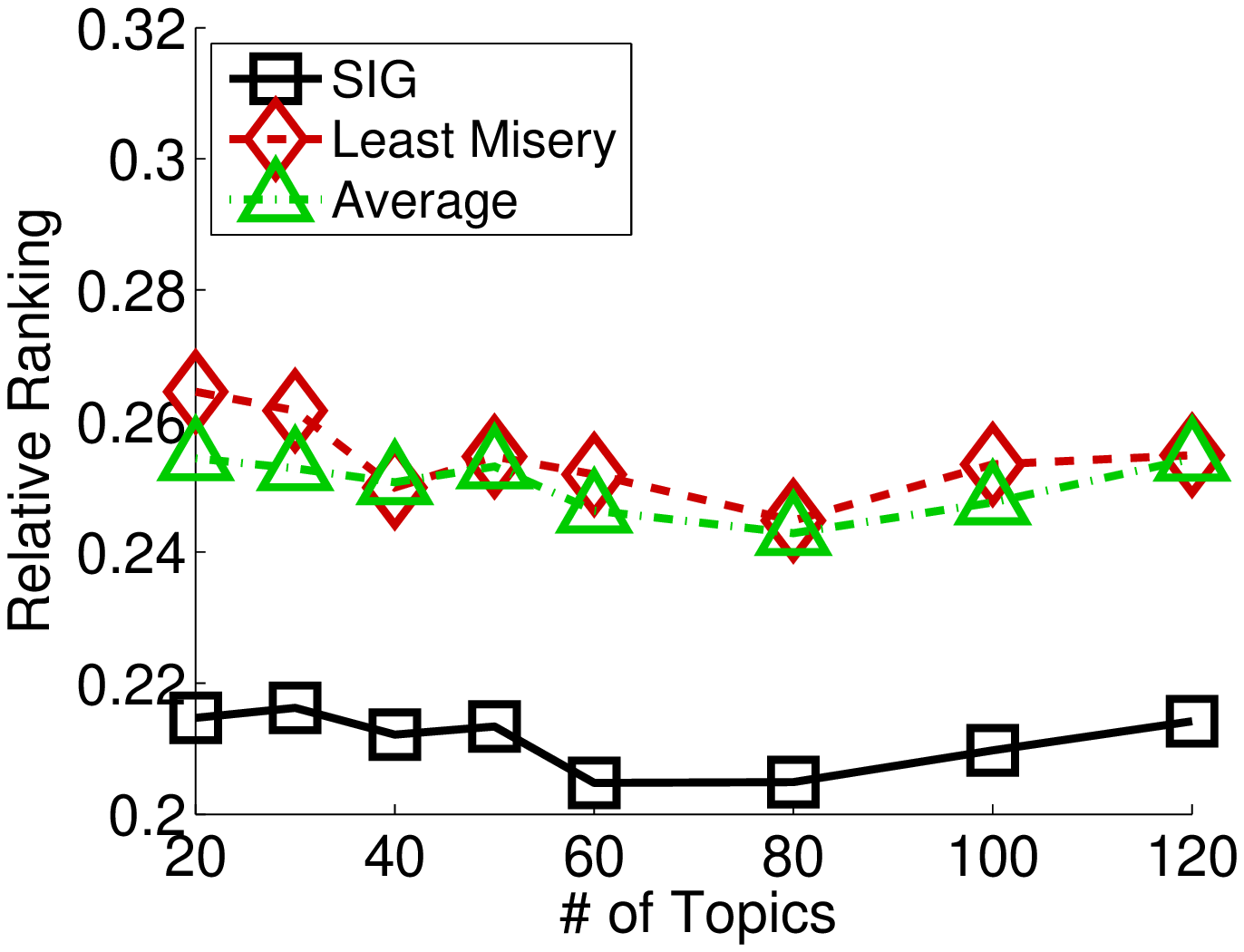}\label{fig:whrrl_group_topic}}
  \caption{Group Recommendation (whrrl.com)}\label{fig:whrrl_group_result}
\end{figure}

In Figure~\ref{fig:whrrl_group_result}, we compare the performance of our SIG group recommendation method with the other two aggregation-based strategies, i.e., Average and Least Misery. The values in Y-axis represent the relative rankings of actual visited places (the lower the better). In Figure~\ref{fig:whrrl_group_size}, we find that SIG outperforms the Average and Least Misery strategies for most of the varied group sizes. However, the larger the group is, the smaller improvement is reached from SIG. This finding implies that for smaller groups, the social influence among group members plays a major role in item selection for the group. However, for larger groups, the group consensus aggregated from individual preferences may dominate the group decision. This finding is consistent with our common experience that in activity planning for a smaller group, one or two influencing members may significantly determine the activity venue. On the other hand, for a large group, the social influence from individuals may be difficult to take effect on the entire group. As a result, the group's common interest dominates. Next, we evaluate the three group recommendation strategies by varying topic size. The result shown in Figure~\ref{fig:whrrl_group_topic} indicates that SIG always outperforms the other two and reach its optimal point when the topic size is configured to around $60$.

\up
\section{Conclusion}\label{sec:conclusion}
This research attempts to explore social influence for item recommendation. We propose a probabilistic generative model, called \emph{unified model}, which naturally unifies the ideas of social influences, collaborative filtering and content-based methods in the recommendation process. To address the issue of hidden social influence in the available datasets, we devise new algorithms to learning the model parameters based on the idea of expectation maximization (EM). Moreover, we provide a Map-Reduce implementation, in addition to a single-machine version, of our EM algorithm to process large-scale datasets. Furthermore, by exploring the social influence quantitatively captured in our models, we develop a social influence based group recommendation algorithm to demonstrate the strength of our proposed models on group recommendation. Finally, we conduct a comprehensive experimental study to evaluate the performance of our proposal for item recommendation to individual users and to groups. Experimental results show that the unified probabilistic generative model proposed in this paper accommodates different factors very well to achieve a superior recommendation performance over other alternatives. Our experimental results also facilitate a better understanding of the social influence between friends in social networks. It is interesting to note that users in whrrl.com are more likely to be influenced by friends than those in last.fm. Finally, our experimental result also confirms that our social influence based algorithm outperforms the state-of-the-art algorithms for group recommendation.   

\section{Acknowledgment}
The authors thank Mu Qiao for insightful discussions.

\newpage

\bibliographystyle{abbrv}

\appendix

\section{EM Algorithm Derivation}
The EM algorithm is a way to find model parameters to achieve local maximum of log-likelihood function (i.e. Equation~(\ref{eq:loglikelihood-cf})). Since direct maximizing $\mathcal{L}(\theta)$ is difficult, EM algorithm applies an iterative method to improve model parameters step by step. Starting from the log-likelihood $\mathcal{L}(\theta)$, we have:
\begin{equation}
\begin{split}
\mathcal{L}(\theta) &= \log\prod_{\langle u,i\rangle \in H}\Pr(u,i|\theta) = \sum_{\langle u,i\rangle \in H}\log\Pr(u,i|\theta)\\
&=  \sum_{\langle u,i\rangle \in H}\log\sum_{z,f}\Pr(u,i,z,f|\theta)\\
&=  \sum_{\langle u,i\rangle \in H}\log\left(\sum_{z,f}\Pr(z,f|u,i,\theta_x)\frac{\Pr(u,i,z,f|\theta)}{\Pr(z,f|u,i,\theta_x)}\right)\\
&\geq \sum_{\langle u,i\rangle \in H}\sum_{z,f}\Pr(z,f|u,i,\theta_x)\log\left(\frac{\Pr(u,i,z,f|\theta)}{\Pr(z,f|u,i,\theta_x)}\right)\\
&\triangleq \mathcal{Q}(\theta|\theta_x)
\end{split}
\end{equation}
Therefore, instead of maximizing $\mathcal{L}(\theta)$, the EM algorithm tries to find model parameters $\theta_{x+1}$ to maximize $\mathcal{Q}(\theta|\theta_x)$. Therefore, we can drop constant terms w.r.t. $\theta$ as
\begin{equation}\label{eq:em_expect}
\begin{split}
\theta_{x+1} &= \arg \max_\theta\{\mathcal{Q}(\theta|\theta_x)\} \\
&= \arg \max_\theta\left\{\sum_{\langle u,i\rangle \in H}\sum_{z,f}\Pr(z,f|u,i,\theta_x)\log\Pr(u,i,z,f|\theta)\right\} \\
&= \arg \max_\theta\left\{\sum_{\langle u,i\rangle \in H}\mathbb{E}_{z,f|u,i,\theta_x}\{\log\Pr(u,i,z,f|\theta)\}\right\}
\end{split}
\end{equation}
Therefore, the EM algorithm consists iterating:
\begin{enumerate}
  \item E-step: Determine the conditional expectation in Equation~(\ref{eq:em_expect}).
  \item M-step: Maximize this expectation with respect to $\theta$.
\end{enumerate}

The E-step needs to find the posterior probabilities in Equation~(\ref{eq:em_expect}), which is computing $\Pr(z,f|u,i,\theta_x)$. Because these probabilities assume model parameters are known as $\theta_x$, we have:
\begin{equation}\label{eq:estep}
\begin{split}
&\Pr(z,f|u,i,\theta_x) \\&= \frac{\Pr(z) \Pr(f|z)\Pr(u|f) \Pr(i|z)}{\sum_{z\in Z}\sum_{f\in F(u)} \Pr(z) \Pr(f|z)\Pr(u|f) \Pr(i|z)}
\end{split}
\end{equation}
where the right hand side of Equation~(\ref{eq:estep}) only consists of the parameters in $\theta_x$.

In the M-step, we need to find model parameters to maximize Equation~(\ref{eq:em_expect}). Firstly, we can break up the term $\log\Pr(u,i,z,f|\theta)$ according to Equation~(\ref{eqn:zifu}) as:
\begin{equation}\label{eqn:log_zifu}
\begin{split}
&\log\Pr(u,f,z,i|\theta)\\
&= \log\Pr(z) + \log \Pr(u|f) + \log\Pr(f|z) + \log\Pr(i|z)
\end{split}
\end{equation}

Plug Equation~(\ref{eqn:log_zifu}) in the the expectation Equation~(\ref{eq:em_expect}) and follow standard calculations, we have:
\begin{equation}\label{eq:separated}
\begin{split}
\theta_{x+1}
&= \arg \max_\theta\\ 
& \sum_{z} \log\Pr(z)\cdot \left(\sum_{\langle u',i'\rangle \in H}\sum_{f'\in F(u')}\Pr(z,f'|u',i')\right)+ \\
&\sum_{u,f} \log \Pr(u|f) \cdot\left(\sum_{\langle u,i'\rangle \in H \wedge f\in F(u)}\sum_{z' \in Z}\Pr(z',f|u,i')\right)+ \\
&\sum_{f,z} \log \Pr(f|z) \cdot\left(\sum_{\langle u',i'\rangle \in H  \wedge f\in F(u')}\Pr(z,f|u',i')\right)+ \\
&\sum_{i,z} \log \Pr(i|z) \cdot\left(\sum_{\langle u',i\rangle \in H}\sum_{f'\in F(u')}\Pr(z,f'|u',i)\right)
\end{split}
\end{equation}

In Equation~\ref{eq:separated}, each model parameters are separated into different inner products. For example, terms related to $\Pr(z)$ is the inner product of $\log\Pr(z) \forall z\in Z$ with corresponding posterior sums. Recall that we always have the probability constrain that $\sum_z \Pr(z) = 1$, to maximize the inner product, $\Pr^+(z)$ should be chosen so that the $\Pr(z)$ vector is at the same ``direction'' as the summed posteriors. In other words, $\Pr^+(z)$ should be proportional to the corresponding summed $\Pr(z,f|u,i)$ \footnote{This vector maximization method is still an approximation, but this approximation is usually good enough and also adopted in \cite{HofmannP99IJCAI}}. Doing the similar maximization to all the model parameters, we can find the $\theta_{x+1} = \{\Pr^+(z), \Pr^+(u|f), \Pr^+(f|z),\Pr^+(i|z)\}$ as:
\begin{subequations}
\begin{align}
\Pr^+(z) & \propto \sum_{\langle u',i'\rangle \in H}\sum_{f' \in F(u')} \Pr(z,f'|u',i')\label{eq:pr_z}\\
\Pr^+(u|f) & \propto \sum_{\langle u,i'\rangle \in H \wedge f\in F(u)}\sum_{z' \in Z} \Pr(z',f|u,i')\label{eq:pr_uf}\\
\Pr^+(f|z) & \propto \sum_{\langle u',i'\rangle \in H \wedge f\in F(u')} \Pr(z,f|u',i') \label{eq:pr_fz}\\
\Pr^+(i|z) & \propto \sum_{\langle u',i\rangle \in H}\sum_{f' \in F(u')} \Pr(z,f'|u',i) \label{eq:pr_iz}
\end{align}\label{eq:pr_mstep}
\end{subequations}

\end{document}